\documentclass[aps,prd,twocolumn,showpacs,superscriptaddress]{revtex4}
\pdfoutput=1
\usepackage{amsmath}
\usepackage{graphicx}

\newcommand{\bastar}{\begin{eqnarray*}}
\newcommand{\eastar}{\end{eqnarray*}}
\newskip\humongous \humongous=0pt plus 1000pt minus 1000pt

\newif\ifdtup

\relax
\newcommand{\W}{{\vec W}}
\newcommand{\n}{{\hat n}}
\newcommand{\hn}{{\hat n}}
\newcommand{\hr}{{\hat r}}
\newcommand{\hD}{{\hat D}}
\newcommand{\bea}{\begin{eqnarray}}
\newcommand{\eea}{\end{eqnarray}}
\newcommand{\A}{{\vec A}}
\newcommand{\valpha}{{\vec \alpha}}

\newcommand{\D}{{\hat D}}
\newcommand{\nn}{\nonumber}
\newcommand{\pro}{\partial}
\newcommand{\Int}{\displaystyle{\int}}
\newcommand{\mn}{{\mu\nu}}
\begin{document}
\title{Finite Energy Electroweak Dyon}
\bigskip

\author{Kyoungtae Kimm}
\affiliation{Faculty of Liberal Education,
Seoul National University, Seoul 151-747, Korea}
\author{J. H. Yoon}
\affiliation{Department of Physics, College of Natural Sciences \\ 
Konkuk University, Seoul 143-701, Korea}
\author{Y. M. Cho}
\email{ymcho7@konkuk.ac.kr}
\affiliation{Administration Building 310-4, Konkuk University, Seoul 143-701, Korea}
\affiliation{School of Physics and Astronomy, Seoul National University,
Seoul 151-742, Korea}

\begin{abstract}
The latest MoEDAL experiment at LHC to detect the electroweak 
monopole makes the theoretical prediction of the monopole mass 
an urgent issue. We discuss three different ways to estimate the mass 
of the electroweak monopole. We first present the dimensional and 
scaling arguments which indicate the monopole mass to be 
around 4 to 10 TeV . To justify this we construct finite energy analytic 
dyon solutions which could be viewed as the regularized Cho-Maison 
dyon, modifying the coupling strength at short distance. Our result 
demonstrates that a genuine electroweak monopole whose 
mass scale is much smaller than the grand unification scale can 
exist, which can actually be detected at the present LHC.
\end{abstract}
\pacs{14.80.Hv, 11.15.Tk, 12.15.-y}
\keywords{mass of electroweak monopole, finite energy 
electroweak monopole, Cho-Maison monopole, Cho-Maison 
dyon}

\maketitle

\section{Introduction}

The recent discover of the Higgs particle at LHC and Tevatron 
has reconfirmed that the electroweak theory of Weinberg and 
Salam provides the true unification of electromagnetic and weak 
interactions \cite{LHC,tev}. Indeed the discovery of the Higgs 
particle has been claimed to be the ``final" test of the standard 
model. This, however, might be a premature claim. The real final 
test should come from the discovery of the electroweak monopole, 
because the standard model predicts it \cite{plb97,yang}. In fact
the existence of the monopole topology in the standard model tells 
that the discovery of the monopole must be the topological test
of the standard model. 

In this sense it is timely that the latest MoEDAL detector 
(``The Magnificient Seventh") at LHC is actively searching for 
such monopole \cite{pin,ijmpa14}. To detect the electroweak monopole 
experimentally, however, it is important to estimate the 
monopole mass in advance. {\it The purpose of this paper is 
to estimate the mass of the electroweak monopole. We show 
that the monopole mass is expected to be around 4 to 7 TeV.} 

Ever since Dirac \cite{Dirac} has introduced the concept of 
the magnetic monopole, the monopoles have remained a fascinating 
subject. The Abelian monopole has been generalized to the non-Abelian 
monopoles by Wu and Yang \cite{wu,prl80} who showed that the pure 
$SU(2)$ gauge theory allows a point-like monopole, and by 't Hooft 
and Polyakov \cite{Hooft,prasad} who have constructed a finite energy 
monopole solution in Georgi-Glashow model as a topological soliton. 
Moreover, the monopole in grand unification has been constructed by
Dokos and Tomaras \cite{dokos}.

In the interesting case of the electroweak theory of Weinberg and 
Salam, however, it has been asserted that there exists no topological 
monopole of physical interest \cite{vach}. The basis for this 
``non-existence theorem'' is, of course, that with the spontaneous 
symmetry breaking the quotient space $SU(2) \times U(1)_Y/U(1)_{\rm em}$ 
allows no non-trivial second homotopy. This has led many people to 
believe that there is no monopole in Weinberg-Salam model. 

This claim, however, has been shown to be false. If the electroweak 
unification of Weinberg and Salam is correct, the standard model 
must have a monopole which generalizes the Dirac monopole. Moreover,
it has been shown that the standard model has a new type of monopole 
and dyon solutions \cite{plb97}. This was based on the observation 
that the Weinberg-Salam model, with the $U(1)_Y$, could be viewed 
as a gauged $CP^1$ model in which the (normalized) Higgs doublet 
plays the role of the $CP^1$ field. So the Weinberg-Salam model 
does have exactly the same nontrivial second homotopy as the 
Georgi-Glashow model which allows topological monopoles.

Once this is understood, one could proceed to construct the desired 
monopole and dyon solutions in the Weinberg-Salam model. Originally 
the electroweak monopole and dyon solutions were obtained by numerical 
integration. But a mathematically rigorous existence proof has been 
established which endorses the numerical results, and the solutions 
are now referred to as Cho-Maison monopole and dyon \cite{yang}.

It should be emphasized that the Cho-Maison monopole is completely 
different from the ``electroweak monopole'' derived from the Nambu's 
electroweak string. In his continued search for the string-like 
objects in physics, Nambu has demonstrated the existence of a rotating 
dumb bell made of the monopole anti-monopole pair connected by the 
neutral string of $Z$-boson flux (actually the $SU(2)$ flux) in 
Weinberg-Salam model \cite{nambu}. Taking advantage of the Nambu's 
pioneering work, others claimed to have discovered another type of 
electroweak monopole, simply by making the string infinitely long 
and moving the anti-monopole to infinity \cite{vacha}. This 
``electroweak monopole'', however, must carry a fractional magnetic 
charge and can not be isolated with finite energy. Moreover, this 
has no spherical symmetry which is manifest in the Cho-Maison 
monopole \cite{plb97}.

The existence of the electroweak monopole makes the experimental
confirmation of the monopole an urgent issue. Till recently the 
experimental effort for the monopole detection has been on the Dirac's 
monopole \cite{cab}. But the electroweak unification of the Maxwell's 
theory requires the modification of the Dirac monopole, and this 
modification changes the Dirac monopole to the Cho-Maison monopole. 
This means that the monopole which should exist in the real world is 
not likely to be the Dirac monopole but the electroweak monopole. 

To detect the electroweak monopole experimentally, it is important 
to estimate the mass of the monopole theoretically. Unfortunately the 
Cho-Maison monopole carries an infinite energy at the classical level, 
so that the monopole mass is not determined. This is because it can 
be viewed as a hybrid between the Dirac monopole and the 'tHooft-Polyakov 
monopole, so that it has a $U(1)_{\rm em}$ point singularity at the center 
even though the $SU(2)$ part is completely regular. 

{\em A priori} there is nothing wrong with this. Classically 
the electron has an infinite electric energy but a finite mass. 
But for the experimental search for the monopole we need a 
solid idea about the monopole mass. In this paper we show 
how to predict the mass of the electroweak monopole. Based 
on the dimensional argument we first show that the monopole 
mass should be of the order of $1/\alpha$ times the W-boson 
mass, or around 10 TeV. To back up this we adopt the scaling 
argument to predict the mass to be around 4 TeV. Finally, 
we show how the quantum correction could regularize
the point singularity of the Cho-Maison dyon, and construct 
finite energy electroweak dyon solutions introducing the effective 
action of the standard model. Our result suggests that the 
electroweak monopole with the mass around 4 to 7 TeV could 
exist, which implies that there is a very good chance that the 
MoEDAL at the present LHC can detect the electroweak 
monopole.

The paper is organized as follows. In Section II we review the 
Cho-Maison dyon for later purpose. In Section III we provide 
two arguments, the dimensional and scaling arguments, which 
indicate that the mass of the electroweak monopole could be 
around 4 to 10 TeV. In Section IV we discuss the Abelian 
decomposition and gauge independent Abelianization of 
Weinberg-Salam model and Georgi-Glashow model to help 
us how to regularize the Cho-Maison monopole. In Section 
V we discuss two different methods to regularize the Cho-Maison 
dyon with the quantum correction which modifies the coupling 
constants at short distance, and construct finite energy dyon 
solutions which support the scaling argument. In Section VI 
we discuss another way to make the Cho-Maison dyon regular, 
by enlarging the gauge group $SU(2)\times U(1)_Y$ to 
$SU(2)\times SU(2)_Y$. Finally in Section VII we discuss 
the physical implications of our results.

\section{Cho-Maison Dyon in Weinberg-Salam Model: A Review}

Before we construct a finite energy dyon solution in the electroweak 
theory we must understand how one can obtain the infinite energy 
Cho-Maison dyon solution first. Let us start with the Lagrangian 
which describes (the bosonic sector of) the Weinberg-Salam theory
\begin{gather}
{\cal L} = -|{\cal D}_{\mu} \phi|^2
 -\frac{\lambda}{2}\big(\phi^\dagger \phi
 -\frac{\mu^2}{\lambda} \big)^2-\frac{1}{4} \vec F_{\mu\nu}^2
 -\frac{1}{4} G_{\mu\nu}^2, \nn\\
{\cal D}_{\mu} \phi=\big(\partial_{\mu} 
-i\frac{g}{2} \vec \tau \cdot \vec A_{\mu} 
- i\frac{g'}{2}B_{\mu}\big) \phi \nn\\
= \big(D_\mu - i\frac{g'}{2}B_{\mu}\big) \phi,
\label{wslag1}
\end{gather}
where $\phi$ is the Higgs doublet, $\vec F_{\mu\nu}$ and $G_{\mu\nu}$
are the gauge field strengths of $SU(2)$ and $U(1)_Y$ with the
potentials $\vec A_{\mu}$ and $B_{\mu}$, and $g$ and $g'$
are the corresponding coupling constants. Notice that $D_{\mu}$ 
describes the covariant derivative  of the $SU(2)$ subgroup only. 
With
\bea
\phi = \dfrac{1}{\sqrt{2}}\rho~\xi,~~~(\xi^{\dagger} \xi = 1),
\eea
where $\rho$ and $\xi$ are the Higgs field and unit doublet, 
we have
\begin{gather}
{\cal L}=-\frac{1}{2}{(\partial_{\mu}\rho)}^2
- \frac{\rho^2}{2} {|{\cal D}_{\mu} \xi |}^2
-\frac{\lambda}{8}\big(\rho^2-\frac{2\mu^2}{\lambda}\big)^2 \nn\\
-\frac{1}{4}{\vec F}_{\mu\nu}^2 -\frac{1}{4} G_{\mu\nu}^2.
\end{gather}
Notice that the $U(1)_Y$ coupling of $\xi$ makes 
the theory a gauge theory of $CP^1$ field \cite{plb97}. 

From (\ref{wslag1}) one has the following equations of motion
\begin{gather}
\pro^2\rho= |{\cal D}_\mu \xi |^2 \rho 
+\frac{\lambda}{2}\big (\rho^2 
- \frac {2\mu^2}{\lambda}\big) \rho, \nn\\
{\cal D}^2 \xi=-2 \dfrac{\partial_\mu \rho}{\rho} 
{\cal D}_\mu \xi +\big[\xi^\dagger {\cal D}^2\xi
+2\dfrac{\partial_\mu \rho}{\rho} 
(\xi^\dagger {\cal D}_\mu \xi)\big] \xi, \nn\\
D_{\mu} \vec F_{\mu\nu}=i \frac{g}{2}\rho^2 \big[\xi^{\dagger} 
\vec \tau( {\cal D}_{\nu} \xi )
-({\cal D}_{\nu} \xi)^{\dagger} \vec \tau \xi \big], \nn\\
\partial_{\mu} G_{\mu\nu}
=i\frac{g'}{2}\rho^2 \big[\xi^{\dagger} ({\cal D}_{\nu} \xi)
- ({\cal D}_{\nu} \xi)^{\dagger} \xi \big].  
\end{gather}
Now we choose the following ansatz in the spherical coordinates
$(t,r,\theta,\varphi)$
\begin{gather}
\rho =\rho(r),  
~~~\xi=i\left(\begin{array}{cc} \sin (\theta/2)~e^{-i\varphi}\\
- \cos(\theta/2) \end{array} \right),   \nn\\
\vec A_{\mu}= \frac{1}{g} A(r)\partial_{\mu}t~\hat r
+\frac{1}{g}(f(r)-1)~\hat r \times \pro_{\mu} \hat r, \nn\\
B_{\mu} =\frac{1}{g'} B(r) \partial_{\mu}t 
-\frac{1}{g'}(1-\cos\theta) \partial_{\mu} \varphi.
\label{ans1}
\end{gather}
Notice that $\xi^{\dagger} \vec \tau ~\xi = -\hat r$. Moreover,
$\A_\mu$ describes the Wu-Yang monopole when $A(r)=f(r)=0$. 
So the ansatz is spherically symmetric. Of course, $\xi$ and 
$B_{\mu}$ have an apparent string singularity along the negative 
$z$-axis, but this singularity is a pure gauge artifact which 
can easily be removed making the $U(1)_Y$ bundle 
non-trivial. So the above ansatz describes a most general 
spherically symmetric ansatz of an electroweak dyon. 

Here we emphasize the importance of the non-trivial nature of 
$U(1)_Y$ gauge symmetry to make the ansatz spherically symmetric. 
Without the extra $U(1)_Y$ the Higgs doublet does not allow a 
spherically symmetric ansatz. This is because the spherical 
symmetry for the gauge field involves the embedding of the radial 
isotropy group $SO(2)$ into the gauge group that requires the 
Higgs field to be invariant under the $U(1)$ subgroup of $SU(2)$. 
This is possible with a Higgs triplet, but not with a Higgs 
doublet \cite{Forg}. In fact, in the absence of the $U(1)_Y$ 
degrees of freedom, the above ansatz describes the $SU(2)$ 
sphaleron which is not spherically symmetric \cite{manton}. 

To see this, one might try to remove the string in $\xi$ with 
the $U(1)$ subgroup of $SU(2)$. But this $U(1)$ will necessarily 
change $\hat r$ and thus violate the spherical symmetry. This 
means that there is no $SU(2)$ gauge transformation which can 
remove the string in $\xi$ and at the same time keeps the spherical 
symmetry intact. The situation changes with the inclusion of the 
$U(1)_Y$ in the standard model, which naturally makes $\xi$ 
a $CP^1$ field \cite{plb97}. This allows the spherical symmetry 
for the Higgs doublet.

To understand the physical content of the ansatz we perform 
the following gauge transformation on (\ref{ans1})
\begin{gather}
\xi \rightarrow U \xi = \left(\begin{array}{cc} 0 \\ 1
\end{array} \right),  \nn\\
U=i\left( \begin{array}{cc}
\cos (\theta/2)& \sin(\theta/2)e^{-i\varphi} \\
-\sin(\theta/2) e^{i\varphi} & \cos(\theta/2)
\end{array}  \right),
\label{gauge}
\end{gather}
and find that in this unitary  gauge we have
\begin{gather}
\hat r \rightarrow \left( \begin{array}{c} 0\\ 0\\
1 \end{array} \right),  \nn\\
\vec A_\mu \rightarrow \frac{1}{g} \left( \begin{array}{c}
-f(r)(\sin\varphi\partial_\mu\theta
+\sin\theta\cos\varphi \partial_\mu\varphi) \\
f(r)(\cos\varphi\partial_\mu \theta
-\sin\theta\sin\varphi\partial_\mu\varphi) \\
A(r)\partial_\mu t -(1-\cos\theta)\partial_\mu\varphi
\end{array} \right).
\label{unitary}
\end{gather}
So introducing the electromagnetic and neutral $Z$-boson potentials 
$A_\mu^{\rm (em)}$ and $Z_\mu$ with the Weinberg angle $\theta_{\rm w}$
\begin{gather}
\left( \begin{array}{cc} A_\mu^{\rm (em)} \\ Z_{\mu}
\end{array} \right)
= \left(\begin{array}{cc}
\cos\theta_{\rm w} & \sin\theta_{\rm w}\\
-\sin\theta_{\rm w} & \cos\theta_{\rm w}
\end{array} \right)
\left( \begin{array}{cc} B_{\mu} \\ A^3_{\mu}
\end{array} \right) \nn\\
= \frac{1}{\sqrt{g^2 + g'^2}} \left(\begin{array}{cc} 
g & g' \\ -g' & g \end{array} \right)
\left( \begin{array}{cc} B_{\mu} \\ A^3_{\mu}
\end{array} \right), 
\label{wein}
\end{gather}
we can express the ansatz (\ref{ans1}) in terms of the physical 
fields
\begin{gather}
W_{\mu} = \frac{1}{\sqrt{2}}(A_\mu^1 + i A_\mu^2)
=\dfrac{i}{g}\frac{f(r)}{\sqrt2}e^{i\varphi}
(\partial_\mu \theta +i \sin\theta \partial_\mu \varphi), \nn\\
A_{\mu}^{\rm (em)} = e\big( \frac{1}{g^2}A(r)
+ \frac{1}{g'^2} B(r) \big) \partial_{\mu}t  \nn\\
-\frac{1}{e}(1-\cos\theta) \partial_{\mu} \varphi,  \nn \\
Z_{\mu} = \frac{e}{gg'}\big(A(r)-B(r)\big) \partial_{\mu}t,
\label{ans2}
\end{gather}
where $W_\mu$ is the $W$-boson and $e$ is the electric charge
\begin{eqnarray*}
e=\frac{gg'}{\sqrt{g^2+g'^2}}=g\sin\theta_{\rm w}=g'\cos\theta_{\rm w}.
\end{eqnarray*}
This clearly shows that the ansatz is for the electroweak dyon.

The spherically symmetric ansatz reduces the equations of 
motion to
\begin{gather}
\ddot{\rho}+\frac{2}{r} \dot{\rho}-\frac{f^2}{2r^2}\rho
=-\frac{1}{4}(A-B)^2\rho +\frac {\lambda}{2}\big(\rho^2
-\frac{2\mu^2}{\lambda}\big)\rho , \nn\\
\ddot{f}-\frac{f^2-1}{r^2}f=\big(\frac{g^2}{4}\rho^2
- A^2\big)f, \nn\\
\ddot{A}+\frac{2}{r}\dot{A}-\frac{2f^2}{r^2}A
=\frac{g^2}{4}\rho^2(A-B), \nn \\
\ddot{B} +\frac{2}{r} \dot{B}
=-\frac{g'^2}{4} \rho^2 (A-B).
\label{cmeq} 
\end{gather}
Obviously this has a trivial solution
\bea
\rho=\rho_0=\sqrt{2\mu^2/\lambda},~~~f=0,
~~~A=B=0,
\eea
which describes the point monopole in Weinberg-Salam model
\bea
A_\mu^{\rm (em)}=-\frac{1}{e}(1-\cos \theta) \partial_\mu \varphi.
\eea
This monopole has two remarkable features. First, this is the 
electroweak generalization of the Dirac's monopole, but not 
the Dirac's monopole. It has the electric charge $4\pi/e$, not 
$2\pi/e$ \cite{plb97}. Second, this monople naturally admits a 
non-trivial dressing of weak bosons. Indeed, with the non-trivial 
dressing, the monopole becomes the Cho-Maison dyon.

To see this let us choose the following boundary condition
\bea
&\rho(0)=0,~~f(0)=1,~~A(0)=0,~~B(0)=b_0, \nn\\
&\rho(\infty)=\rho_0,~f(\infty)=0,~A(\infty)=B(\infty)=A_0.
\label{bc0}
\eea
Then we can show that the equation (\ref{cmeq}) admits 
a family of solutions labeled by the real parameter $A_0$ 
lying in the range \cite{plb97,yang} 
\bea
0 \leq A_0 < {\rm min} ~\Big(e\rho_0,~\frac{g}{2}\rho_0\Big).
\label{boundA} 
\eea 
In this case all four functions
$f(r),~\rho(r),~A(r)$, and $B(r)$ must be positive for $r>0$, and
$A(r)/g^2+B(r)/g'^2$ and $B(r)$ become increasing functions of
$r$. So we have $0 \leq b_0 \leq A_0$. Furthermore, we have
$B(r)\ge A(r)\ge 0$ for all range, and $B(r)$ must approach to
$A(r)$ with an exponential damping. Notice that, with the
experimental fact $\sin^2\theta_{\rm w}=0.2312$, (\ref{boundA})
can be written as $0 \leq A_0 < e\rho_0$.

With the boundary condition (\ref{bc0}) we can integrate
(\ref{cmeq}). For example, with $A=B=0$, we have the 
Cho-Maison monopole. In general, with $A_0\ne0$, we find 
the Cho-Maison dyon \cite{plb97}.

Near the origin the dyon solution has the following behavior,
\bea
&\rho \simeq \alpha_1 r^{\delta_-},
~~~~~f \simeq 1+ \beta_1  r^2,  \nonumber \\
&A \simeq a_1 r,~~~~~B \simeq b_0 + b_1 r^{2\delta_+},
\label{origin}
\eea
where $\delta_{\pm} =(\sqrt{3} \pm 1)/2$.
Asymptotically it has the following behavior,
\bea
&\rho \simeq \rho_0 +\rho_1\dfrac{\exp(-\sqrt{2}\mu r)}{r}, 
~~~f \simeq  f_1 \exp(-\omega  r),  \nn\\
&A \simeq A_0 +\dfrac{A_1}{r},
~~~B \simeq A +B_1 \dfrac{\exp(-\nu r)}{r}, 
\label{infty}
\eea
where $\omega=\sqrt{(g\rho_0)^2/4 -A_0^2}$,
and $\nu=\sqrt{(g^2 +g'^2)}\rho_0/2$.
The physical meaning of the asymptotic behavior
must be clear. Obviously $\rho$, $f$, and $A-B$ represent
the Higgs boson, $W$-boson, and $Z$-boson whose masses
are given by $M_H=\sqrt{2}\mu=\sqrt{\lambda}\rho_0$,
$M_W=g\rho_0/2$, and $M_Z=\sqrt{g^2+g'^2}\rho_0/2$. 

So (\ref{infty}) tells that $M_H$, $\sqrt{1-(A_0/M_W)^2}~M_W$,
and $M_Z$ determine the exponential damping of the Higgs boson, 
$W$-boson, and $Z$-boson to their vacuum expectation values 
asymptotically. Notice that it is $\sqrt{1-(A_0/M_W)^2}~M_W$, 
but not $M_W$, which determines the exponential damping of the 
$W$-boson. This tells that the electric potential of the dyon
slows down the exponential damping of the $W$-boson, which is 
reasonable.

The dyon has the following electromagnetic charges
\bea
&q_e=-\dfrac{8\pi}{e}\sin^2\theta_{\rm w} \Int_0^\infty f^2 A dr 
=\dfrac{4\pi}{e} A_1, \nn\\
&q_m = \dfrac{4\pi}{e}. 
\label{eq:Charge}
\end{eqnarray}
Also, the asymptotic condition (\ref{infty}) assures
that the dyon does not carry any neutral charge,
\begin{eqnarray}
&Z_e =-\dfrac{4\pi e}{gg'}\big[ r^2 (\dot{A}-\dot{B})\big]
\Big|_{r=\infty} =0,\nn\\
&Z_m = 0.
\label{neutral}
\end{eqnarray}
Furthermore, notice that the dyon equation (\ref{cmeq})
is invariant under the reflection
\bea
A \rightarrow -A,~~~~~~B\rightarrow -B.
\label{ref}
\eea
This means that, for a given magnetic charge,
there are always two dyon solutions
which carry opposite electric charges $\pm q_e$.
Clearly the signature of
the electric charge of the dyon is determined by
the signature of the boundary value $A_0$.

We can also have the anti-monopole or in general anti-dyon 
solution, the charge conjugate state of the dyon, which 
has the magnetic charge $q_m=-4\pi/e$ with the following 
ansatz
\begin{gather}
\rho' =\rho(r),  
~~~\xi'=-i\left(\begin{array}{cc} \sin (\theta/2)~e^{+i\varphi}\\
- \cos(\theta/2) \end{array} \right),   \nn\\
\vec A'_{\mu}= -\frac{1}{g} A(r)\partial_{\mu}t~\hat r'
+\frac{1}{g}(f(r)-1)~\hat r' \times \pro_{\mu} \hat r', \nn\\
B'_{\mu} =-\frac{1}{g'} B(r) \partial_{\mu}t 
+\frac{1}{g'}(1-\cos\theta) \partial_{\mu} \varphi,  \nn\\
\hat r'=-\xi'^{\dagger} \vec \tau ~\xi'
=(\sin \theta \cos \phi,-\sin \theta \sin \phi,\cos \theta).
\label{antid1}
\end{gather}
Notice that the ansatz is basically the complex conjugation 
of the dyon ansatz. 

To understand the meaning of the anti-dyon ansatz notice that 
in the unitary gauge 
\begin{gather}
\xi' \rightarrow U' \xi' = \left(\begin{array}{cc} 0 \\ 1
\end{array} \right),
\nn\\
U'=-i\left( \begin{array}{cc}
\cos (\theta/2)& \sin(\theta/2)e^{i\varphi} \\
-\sin(\theta/2) e^{-i\varphi} & \cos(\theta/2)
\end{array} \right),
\end{gather}
we have
\begin{gather}
\A'_\mu \rightarrow \frac{1}{g} \left( \begin{array}{c}
f(r)(\sin\varphi\partial_\mu\theta
+\sin\theta\cos\varphi \partial_\mu\varphi) \\
f(r)(\cos\varphi\partial_\mu \theta
-\sin\theta\sin\varphi\partial_\mu\varphi) \\
-A(r)\partial_\mu t +(1-\cos\theta)\partial_\mu\varphi
\end{array} \right).
\label{unitary}
\end{gather}
So in terms of the physical fields the ansatz (\ref{antid1})
is expressed by 
\bea
&W'_{\mu}=\dfrac{i}{g}\dfrac{f(r)}{\sqrt2}e^{-i\varphi}
(\partial_\mu \theta -i \sin\theta \partial_\mu \varphi)
=-W_{\mu}^*, \nn\\
&A_{\mu}^{ \rm (em)} = -e\big( \dfrac{1}{g^2}A(r)
+ \dfrac{1}{g'^2} B(r) \big) \partial_{\mu}t  \nn\\
&+\dfrac{1}{e}(1-\cos\theta) \partial_{\mu} \varphi,  \nn \\
&Z'_{\mu} = -\dfrac{e}{gg'}\big(A(r)-B(r)\big) \partial_{\mu}t
=-Z_\mu.
\label{antid2}
\eea
This clearly shows that the the electric and magnetic charges
of the ansatz (\ref{antid1}) are the opposite of the dyon ansatz, 
which confirms that the ansatz indeed describes the anti-dyon.

With the ansatz (\ref{antid1}) we have exactly the same 
equation (\ref{cmeq}) for the anti-dyon. This assures 
that the standard model has the anti-dyon as well as 
the dyon.

The above discussion tells that the W and Z boson part of 
the anti-dyon solution is basically the complex conjugation 
of the dyon solution. This, of course, is natural from 
the physical point of view. On the other hand there is 
one minor point to be clarified here. Since the topological 
charge of the monopole is given by the second homotopy 
defined by $\hat r=-\xi^\dagger \vec \tau \xi$, one might 
expect that $\hat r'$ defined by the anti-dyon ansats 
$\xi'=\xi^*$ must be $-\hat r$. But this is not so, and 
we have to explain why. 

To understand this notice tha we can change $\hat r'$ 
to $-\hat r$ by a SU(2) gauge transformation, by the 
$\pi$-rotation along the y-axis. With this gauge 
transformation the ansatz (\ref{antid1}) changes to 
\begin{gather}
\xi' \rightarrow i\left(\begin{array}{cc} \cos(\theta/2) \\
\sin (\theta/2)~e^{+i\varphi} \end{array} \right),
~~~\hat r' \rightarrow -\hat r,  \nn\\
\vec A_{\mu} \rightarrow  -\frac{1}{g} A(r)\partial_{\mu}t~\hat r
+\frac{1}{g}(f(r)-1)~\hat r \times \pro_{\mu} \hat r.
\label{antid3}
\end{gather}
This tells that the monopole topology defined by $\hat r'$ is 
the same as that of $\hat r$. 

Since the Cho-Maison solution is obtained numerically one 
might like to have a mathematically rigorous existence proof
of the Cho-Maison dyon. The existence proof is non-trivial, 
because the equation of motion (\ref{cmeq}) is not the 
Euler-Lagrange equation of the positive definite energy 
(\ref{cme}), but that of the indefinite action
\begin{gather}
{\cal L}=-4\pi \int\limits_0^\infty dr 
\bigg\{\frac{1}{2}(r\dot\rho)^2 
+\frac{\lambda r^2}{8}\big(\rho^2-\rho_0^2\big)^2 \nn\\
+\frac{1}{4} f^2\rho^2+ \frac1{g^2} \big(\dot f^2
-\frac{1}{2}(r\dot A)^2- f^2 A^2 \big)-\frac{1}{2g'^2}(r\dot B)^2  \nn\\
-\frac{r^2}{8} (B-A)^2 \rho^2 +\frac{1}{2 r^2}\big(\frac{1}{g'^2}
+\frac1{g^2}(f^2-1)^2\big) \bigg\}.
\label{cmlag}
\end{gather}
Fortunately the existence proof has been established by 
Yang \cite{yang}.

Before we leave this section it is worth to re-address the 
important question again: Does the standard model predict 
the monopole? Notice that the Dirac monopole in electrodynamics 
is optional: It can exist only when the $U(1)_{\rm em}$ is 
non-trivial, but there is no reason why this has to be so. 
If so, why can't the electroweak monopole be optional?

As we have pointed out, the non-trivial $U(1)_Y$ is crucial 
for the existence of the monopole in the standard model. 
So the question here is why the $U(1)_Y$ must be non-trivial.  
To see why, notice that in the standard model $U(1)_{\rm em}$
comes from two $U(1)$, the $U(1)$ subgroup of $SU(2)$ and 
$U(1)_Y$, and it is well known that the $U(1)$ subgroup of $SU(2)$ 
is non-trivial. Now, to obtain the electroweak monopole we have 
to make the linear combination of two monopoles, that of the 
$U(1)$ subgroup of $SU(2)$ and $U(1)_Y$. This must be clear 
from (\ref{wein}).

In this case the mathematical consistency requires the two 
potentials $A_\mu^3$ and $B_\mu$ (and two $U(1)$) to have 
the same structure, in particular the same topology. But we 
already know that $A_\mu^3$ is non-trivial. So $B_\mu$,  
and the corresponding $U(1)_Y$, has to be non-trivial. In 
other words, requiring $U(1)_Y$ to be trivial is inconsistent
(i.e., in contradiction with the self-consistency) in the 
standard model. This tells that, unlike the Maxwell's theory, 
the $U(1)_{\rm em}$ in the standard model must be non-trivial. 
This assures that the standard model must have the monopole.

But ultimately this question has to be answered by the experiment. 
So the discovery of the monopole must be the topological test 
of the standard model, which has never been done before. This 
is why MoEDAL is so important. 

\section{Mass of the Electroweak Monopole}

To detect the electroweak monopole experimentally, we have to 
have a firm idea on the mass of the monopole. Unfortunately, 
at the classical level we can not estimate the mass of the 
Cho-Maison monopole, because it has a point singularity at the 
center which makes the total energy infinite.

Indeed the ansatz (\ref{ans1}) gives the following energy 
\begin{gather}
E=E_0 +E_1,  \nonumber \\
E_0=4\pi\int_0^\infty \frac{dr}{2 r^2}
\bigg\{\frac{1}{g'^2}+ \frac1{g^2}(f^2-1)^2\bigg\}, \nn\\
E_1=4\pi \int_0^\infty dr \bigg\{\frac12 (r\dot\rho)^2
+\frac1{g^2} \big(\dot f^2 +\frac{1}{2}(r\dot A)^2  \nn\\
+ f^2 A^2 \big)+\frac{1}{2g'^2}(r\dot B)^2 
+\frac{\lambda r^2}{8}\big(\rho^2-\rho_0^2 \big)^2 \nn\\
+\frac14 f^2\rho^2
+\frac{r^2}{8} (B-A)^2 \rho^2 \bigg\}.
\label{cme}
\end{gather}
The boundary condition (\ref{bc0}) guarantees that
$E_1$ is finite. As for $E_0$ we can minimize it with 
the boundary condition $f(0)=1$, but even with this 
$E_0$ becomes infinite. Of course the origin of this
infinite energy is obvious, which is precisely due to 
the magnetic singularity of $B_\mu$ at the origin.
This means that one can not predict the mass of dyon.
Physically it remains arbitrary.

To estimate of the monopole mass theoretically, we have to 
regularize the point singularity of the Cho-Maison dyon. One 
might try to do that introducing the gravitational interaction, 
in which case the mass is fixed by the asymptotic behavior of 
the gravitational potential. But the magnetic charge of the 
monopole is not likely to change the character of the singularity, 
so that asymptotically the leading order of the gravitational 
potential becomes of the Reissner-Nordstrom type \cite{bais}. 
This implies the gravitational interaction may not help us to 
estimate the monopole mass. 

To make the the energy of the Cho-Maison monopole finite, 
notice that the origin of the infinite energy is the first 
term $1/g'^2$ in $E_0$ in (\ref{cme}). A simple way to make 
this term finite is to introduce a UV-cutoff which removes 
this divergence. This type of cutoff could naturally come 
from the quantum correction of the coupling constants. In 
fact, since the quantum correction changes $g'$ to the 
running coupling $\bar g'$, $E_0$ can become finite 
if $\bar g'$ diverges at short distance. 

We will discuss how such quantum correction could take place 
later, but before doing that we present two arguments, the 
dimsnsional argument and the scaling argument, which 
could give us a rough estimate of the monopole mass. 

\subsection{Dimensional argument}  

To have the order estimate of the monopole mass it is important 
to realize that, roughly speaking, the monopole mass comes from 
the Higgs mechanism which generates the mass to the W-boson. 
This can easily be seen in the 'tHooft-Polyakov monopole in
Georgi-Glashow model
\bea
&{\cal L}_{GG} =-\dfrac{1}{4} \vec{F}_{\mu\nu}^2
-\dfrac{1}{2}(D_\mu \vec{\Phi} )^2-\dfrac{\lambda}{4} 
\big(\vec \Phi^2-\dfrac{\mu^2}{\lambda}\big)^2,
\label{ggl}
\eea
where $\vec \Phi$ is the Higgs triplet. Here the monopole ansatz 
is given by
\bea
&\vec \Phi=\rho~\hat r,~~~\A_\mu= \vec C_\mu+ \W_\mu,  \nn\\
&\vec C_\mu= -\dfrac{1}{g} \hat r \times \pro_\mu \hat r,
~~~\W_\mu=-f \vec C_\mu,
\label{wu}
\eea
where $\vec C_\mu$ represents the Wu-Yang monopole 
potential \cite{wu,prd80}. Notice that the W-boson
part of the monopole is given by the Wu-Yang potential, 
except for the overall amplitude $f$. 

With this we clearly have
\bea
|D_\mu \Phi|^2=(\pro_\mu \rho)^2
+ g^2 \rho^2 f^2 (\vec C_\mu)^2. 
\eea
So, when the Higgs field has a non-vanishing vacuum expectation 
value, $\vec C_\mu$ acquires a mass (with $f\simeq 1$). This, 
of course, is the Higgs mechanism which generates the W-boson
mass. The only difference is that here the W-boson is expressed 
by the Wu-Yang potential and the Higgs coupling becomes magnetic 
($\vec C_\mu$ contains the extra factor $1/g$). 

Similar mechanism works for the Weinberg-Salam model. Here again 
$\A_\mu$ (with $A=B=0$) of the asnatz (\ref{ans1}) is identical to 
(\ref{wu}), and we have
\begin{gather}
D_\mu \xi=i\big(g f~\vec C_\mu+(1-\cos \theta) \pro_\mu \phi~\hr \big)
\cdot \frac{\vec \tau}{2}~\xi,  \nn\\
|{\cal D}_\mu \xi|^2=|D_\mu \xi|^2- |\xi^\dagger D_\mu \xi|^2  \nn\\
-(\xi^\dagger D_\mu\xi-i \frac{g'}{2} B_\mu)^2
=\frac14 g^2 f^2 (\vec C_\mu)^2, \nn\\
|{\cal D}_\mu \phi|^2=\frac12 (\pro_\mu \rho)^2
+\frac12 \rho^2 |{\cal D}_\mu \xi|^2 \nn\\
=\frac12 (\pro_\mu \rho)^2+\frac18 g^2 \rho^2 f^2 (\vec C_\mu)^2. 
\end{gather}
This (with $f\simeq 1$) tells that the electroweak monopole 
acquires mass through the Higgs mechanism which generates mass 
to the W-boson.

Once this is understood, we can use the dimensional argument 
to predict the monopole energy. Since the monopole mass term 
in the Lagrangian contributes to the monopole energy in the 
classical solution we may expect
\bea
E \simeq C \times \dfrac{4\pi}{e^2} M_W,
~~~C\simeq 1.
\label{mmass}
\eea 
This implies that the monopole mass should be about $1/\alpha$ 
times bigger than the electroweak scale, around 10 TeV. But this 
is the order estimate. Now we have to know how to estimate $C$.

\subsection{Scaling argument}  

We can use the Derrick's scaling argument to estimate the 
constant $C$ in (\ref{mmass}), assuming the existence of 
a finite energy monopole solution. If a finite energy monopole 
does exist, the action principle tells that it should be stable 
under the rescaling of its field configuration. So consider such 
a monopole configuration and let
\begin{gather}
K_A = \Int d^3x ~\dfrac{1}{4} \vec{F}_{ij}^2,
\quad K_B=\Int d^3x ~\dfrac{1}{4} B_{ij}^2  \nn\\
K_\phi=\Int d^3 x ~|{\cal D}_i \phi|^2, \nn\\
V_\phi=\Int d^3x ~\dfrac{\lambda}{2}\big( |\phi|^2
-\dfrac{\mu^2}{\lambda} \big)^2. 
\end{gather}
With the ansatz (\ref{ans1}) we have (with $A=B=0$)
\begin{gather}
K_A= \frac{4\pi}{g^2} \int_0^\infty 
\Big\{\dot{f}^2 + \frac{(f^2-1)^2}{2r^2} \Big\} dr, \nn\\
K_B=\frac{2\pi}{g'^2}\int\limits_0^\infty \frac{1}{r^2}dr,
~~~K_\phi= 2\pi \int_0^\infty (r\dot{\rho})^2 dr,  \nn \\
V_\phi=\frac{\pi}{2} \int_0^\infty \lambda r^2 
\big(\rho^2 -\rho_0^2 \big)^2 dr.
\end{gather}
Notice that $K_B$ makes the monopole energy infinite. 

Now, consider the spatial scale transformation 
\bea
\vec x \longrightarrow \lambda \vec x.
\label{scale}
\eea
Under this we have 
\bea
&\A_k(\vec x) \rightarrow \lambda \A_k(\lambda \vec x),
~~~B_k(\vec x)\rightarrow \lambda B_k(\lambda \vec x), \nn\\
&\phi (\vec x) \rightarrow \phi (\lambda \vec x),
\eea
so that 
\bea
&K_A \longrightarrow \lambda K_A, 
~~~K_B \longrightarrow \lambda K_B,  \nn\\
&K_\phi \longrightarrow \lambda^{-1} K_\phi,
~~~V_\phi \longrightarrow \lambda^{-3} V_\phi.
\eea
With this we have the following requirement for the stable 
monopole configuration 
\bea
K_A+K_B=K_\phi+3V_\phi.
\label{derrick}
\eea
From this we can estimate the finite value of $K_B$. 

Now, for the Cho-Maison monopole we have (with 
$M_W \simeq 80.4~{\rm GeV}$, $M_H \simeq 125~{\rm GeV}$, 
and $\sin^2\theta_{\rm w}=0.2312$) 
\bea
&K_A \simeq 0.1904 \times\dfrac{4\pi}{e^2}{M_W},
~~~K_\phi \simeq 0.1577 \times\dfrac{4\pi}{e^2}{M_W},  \nn \\
&V_\phi \simeq 0.0111 \times\dfrac{4\pi}{e^2}{M_W}.
\eea
This, with (\ref{derrick}), tells that
\bea
K_B \simeq 0.0006 \times \dfrac{4\pi}{e^2} M_W.
\eea
From this we estimate the energy of the monopole to be
\bea
E \simeq 0.3598 \times \dfrac{4\pi}{e^2} M_W \simeq 3.96~{\rm TeV}.
\eea
This strongly endorses the dimensional argument. In particular, this  
tells that the electroweak monopole of mass around a few TeV could 
be possible. 

The important question now is to show how the quantum correction 
could actually make the energy of the Cho-Maison monopole finite. 
To do that we have to understand the structure of the electroweak 
theory, in particular the Abelian decomposition of the electroweak 
theory. So we review the gauge independent Abelian decomposition 
of the standard model first.

\section{Abelian Decomposition of the Electroweak Theory}

Consider the Yang-Mills theory
\begin{eqnarray}
{\cal L}_{YM} =-\dfrac{1}{4} \vec F_{\mu\nu}^2.
\end{eqnarray}
A best way to make the Abelian decomposition is to introduce 
a unit $SU(2)$ triplet $\hat n$ which selects the Abelian 
direction at each space-time point, and impose the isometry 
on the gauge potential which determines the restricted potential 
$\hat A_\mu$ \cite{prd80,prl81}
\bea
&D_\mu \hn=0,  \nn\\
&\vec A_\mu\rightarrow \hat A_\mu 
=A_\mu \n -\dfrac{1}{g} \n\times\pro_\mu\n
=A_\mu \n+\vec C_\mu, \nn\\
&A_\mu=\n \cdot \A_\mu,
~~~\vec C_\mu=-\dfrac{1}{g} \n\times\pro_\mu\n .
\label{chocon}
\eea 
Notice that the restricted potential is precisely the connection 
which leaves $\n$ invariant under parallel transport. The restricted 
potential is called Cho connection or Cho-Duan-Ge (CDG) 
connection \cite{fadd,shab,zucc}.

With this we obtain the gauge independent Abelian decomposition
of the $SU(2)$ gauge potential adding the valence potential 
$\vec W_\mu$ which was excluded by the isometry \cite{prd80,prl81}
\begin{gather} 
\vec{A}_\mu = \hat A_\mu + \W_\mu, 
~~~(\hat{n}\cdot\vec{W}_\mu=0).
\label{chodecom}
\end{gather}
The Abelian decomposition has recently been referred 
to as Cho (also Cho-Duan-Ge or Cho-Faddeev-Niemi) 
decomposition \cite{fadd,shab,zucc}.

Under the infinitesimal gauge transformation 
\bea 
\delta \n = - \vec \alpha \times \n, 
~~~~\delta \A_\mu = \frac{1}{g}  D_\mu \vec \alpha, 
\eea
we have
\begin{gather}
\delta A_\mu = \frac{1}{g} \n \cdot \pro_\mu \valpha,
\quad \delta \hat A_\mu = \frac{1}{g} \D_\mu \valpha, \nn\\
\delta \W_\mu = -\valpha \times \W_\mu.
\label{gt1}
\end{gather}
This tells that $\hat A_\mu$ by itself describes an $SU(2)$
connection which enjoys the full $SU(2)$ gauge degrees of freedom.
Furthermore the valence potential $\vec W_\mu$ forms a gauge
covariant vector field under the gauge transformation. But
what is really remarkable is that the decomposition is gauge
independent. Once $\hat n$ is chosen, the decomposition follows
automatically, regardless of the choice of gauge. 

Notice that $\hat{A}_\mu$ has a dual structure,
\begin{gather}
\hat{F}_{\mu\nu} 
= \partial_\mu \hat A_\nu-\partial_\nu \hat A_\mu
+ g \hat A_\mu \times \hat A_\nu 
= (F_{\mu\nu}+ H_{\mu\nu})\hat{n}, \nonumber \\
F_{\mu\nu} =\partial_\mu A_\nu-\partial_\nu A_\mu, \nn\\
H_{\mu\nu} = -\frac{1}{g} \hat{n}\cdot(\partial_\mu
\hat{n}\times\partial_\nu\hat{n}).
\end{gather}
Moreover, $H_{\mu \nu}$ always admits the potential because 
it satisfies the Bianchi identity. In fact, replacing $\hn$ 
with a $CP^1$ field $\xi$ (with $\hn=-\xi^{\dagger} \vec \tau ~\xi$) 
we have
\bea
&H_{\mu\nu} = \partial_\mu \tilde C_\nu-\partial_\nu \tilde C_\mu 
= \dfrac{2i}{g} (\partial_\mu \xi^{\dagger}
\partial_\nu \xi - \partial_\nu \xi^{\dagger} \partial_\mu \xi), \nn\\
&\tilde C_\mu = \dfrac{2i}{g} \xi^{\dagger} \partial_\mu \xi
=\dfrac{i}{g} \big(\xi^{\dagger} \partial_\mu \xi
- \partial_\mu \xi^{\dagger} \xi \big).
\eea
Of course $\tilde C_\mu$ is determined uniquely up to the $U(1)$ 
gauge freedom which leaves $\n$ invariant. To understand the 
meaning of $\tilde C_\mu$, notice that with $\hat n=\hat r$ we 
have 
\bea
\tilde C_\mu = \frac{1}{g} (1- \rm cos~\theta) \partial_\mu \varphi.
\eea
This is nothing but the Abelian monopole potential, and the 
corresponding non-Abelian monopole potential is given by
the Wu-Yang monopole potential $\vec C_\mu$ \cite{wu,prl80}. 
This justifies us to call $A_\mu$ and $\tilde C_\mu$ the electric 
and magnetic potential.

The above analysis tells that $\hat{A}_\mu$ retains all essential 
topological characteristics of the original non-Abelian potential. 
First, $\hat{n}$ defines $\pi_2(S^2)$ which describes the non-Abelian 
monopoles. Second, it characterizes the Hopf invariant 
$\pi_3(S^2)\simeq\pi_3(S^3)$ which describes the topologically 
distinct vacua \cite{bpst,plb79}. Moreover, it provides the gauge 
independent separation of the monopole field from the generic 
non-Abelian gauge potential. 

With the decomposition (\ref{chodecom}), we have
\bea
\vec{F}_{\mu\nu} 
=\hat F_{\mu \nu} + \D _\mu \W_\nu - \D_\nu
\W_\mu + g\W_\mu \times \W_\nu, 
\eea 
so that the Yang-Mills Lagrangian is expressed as
\begin{gather}
{\cal L}_{YM} =-\frac{1}{4} {\hat F}_{\mu\nu}^2 
-\frac{1}{4}(\D_\mu\W_\nu-\D_\nu\W_\mu)^2 \nn\\
-\frac{g}{2}{\hat F}_{\mu\nu} \cdot (\W_\mu \times \W_\nu)
-\frac{g^2}{4} (\W_\mu \times \W_\nu)^2. 
\end{gather} 
This shows that the Yang-Mills theory can be viewed as 
a restricted gauge theory made of the restricted potential, 
which has the valence gluons as its source \cite{prd80,prl81}. 

An important advantage of the decomposition (\ref{chodecom}) is 
that it can actually Abelianize (or more precisely ``dualize'') 
the non-Abelian gauge theory gauge independently \cite{prd80,prl81}. 
To see this let$(\n_1,~\n_2,~\n)$ be a right-handed orthonormal 
basis of $SU(2)$ space and let
\begin{gather}
\vec{W}_\mu =W^1_\mu ~\hat{n}_1 + W^2_\mu ~\hat{n}_2,
\nn\\
(W^1_\mu = \hat {n}_1 \cdot \vec W_\mu,~~~W^2_\mu =
\hat {n}_2 \cdot \vec W_\mu).         
\nonumber
\end{gather}
With this we have
\bea
&\hat{D}_\mu \vec{W}_\nu =\Big[\partial_\mu W^1_\nu
-g(A_\mu+ \tilde C_\mu)W^2_\nu \Big]\hat n_1  \nn\\
&+\Big[\partial_\mu W^2_\nu
+ g (A_\mu+ \tilde C_\mu)W^1_\nu \Big]\hat{n}_2,
\eea
so that with
\bea
{\cal A}_\mu = A_\mu+ \tilde C_\mu,
\quad
W_\mu = \frac{1}{\sqrt{2}} ( W^1_\mu + i W^2_\mu ), \nn
\eea
we can express the Lagrangian explicitly in terms of the dual
potential ${\cal A}_\mu$ and the complex vector field $W_\mu$,
\begin{eqnarray} 
\label{eq:Abelian}
{\cal L}_{YM} = -\frac{1}{4} {\cal F}_{\mu\nu}^2
-\frac{1}{2}|\hat{D}_\mu{W}_\nu-\hat{D}_\nu{W}_\mu|^2 \nn\\
+ ig {\cal F}_{\mu\nu} W_\mu^*W_\nu
+ \frac{g^2}{4}(W_\mu^* W_\nu - W_\nu^* W_\mu)^2,
\end{eqnarray}
where ${\cal F}_{\mu\nu} = F_{\mu\nu} + H_{\mu\nu}$ and 
$\hat{D}_\mu=\partial_\mu + ig {\cal A}_\mu$. This shows 
that we can indeed Abelianize the non-Abelian theory 
with our decomposition.

Notice that in the Abelian formalism the Abelian potential 
${\cal A}_\mu$ has the extra magnetic potential $\tilde C_\mu$. 
In other words, it is given by the sum of the electric and magnetic 
potentials $A_\mu+\tilde C_\mu$. Clearly $\tilde C_\mu$ 
represents the topological degrees of the non-Abelian 
symmetry which does not show up in the naive Abelianization that 
one obtains by fixing the gauge \cite{prd80,prl81}.

Furthermore, this Abelianization is gauge independent, 
because here we have never fixed the gauge to obtain 
this Abelian formalism. So one might ask how the non-Abelian 
gauge symmetry is realized in this Abelian formalism. To 
discuss this let
\begin{gather}
\vec \alpha = \alpha_1~\hn_1 + \alpha_2~\hn_2 + \theta~\hat n,
\quad
\alpha = \frac{1}{\sqrt 2} (\alpha_1 + i ~\alpha_2), 
\nn\\
\vec C_\mu = - \frac {1}{g} \hn \times \partial_\mu \hn
= - C^1_\mu \hn_1 - C^2_\mu \hn_2,
\nn\\
C_\mu = \frac{1}{\sqrt 2} (C^1_\mu + i ~ C^2_\mu).
\end{gather}
Certainly the Lagrangian (\ref{eq:Abelian}) is invariant under 
the active (classical) gauge transformation (\ref{gt1}) described 
by
\begin{gather}
\delta A_\mu = \frac{1}{g} \partial_\mu \theta 
- i (C_\mu^* \alpha - C_\mu \alpha^*),
\nn\\
\delta \tilde C_\mu = - \delta A_\mu,
\quad 
\delta W_\mu = 0.
\label{eq:active}
\end{gather}
But it has another gauge invariance, the invariance under the 
following passive (quantum) gauge transformation
\begin{gather} 
\delta A_\mu = \frac{1}{g} \partial_\mu \theta 
-i (W_\mu^* \alpha - W_\mu \alpha^*),
\nn\\
\delta \tilde C_\mu = 0,
\quad
\delta W_\mu = \frac{1}{g} \hD_\mu \alpha - i \theta W_\mu.
\label{eq:passive}
\end{gather}
Clearly this passive gauge transformation assures the desired
non-Abelian gauge symmetry for the Abelian formalism.
This tells that the Abelian theory not only retains
the original gauge symmetry, but actually has an enlarged (both
active and passive) gauge symmetries.

The reason for this extra (quantum) gauge symmetry is that 
the Abelian decomposition automatically put the theory in 
the background field formalism which doubles the gauge 
symmetry \cite{dewitt}. This is because in this decomposition 
we can view the restricted and valence potentials as the 
classical and quantum potentials, so that we have freedom to 
assign the gauge symmetry either to the classical field 
or to the quantum field. This is why we have the extra gauge 
symmetry.

The Abelian decomposition has played a crucial role in QCD 
to demonstrate the Abelian dominance and the monopole condensation 
in color confinement \cite{prd00,prd13,kondo}. This is because 
it separates not only the Abelian potential but also the monopole 
potential gauge independently. 

Now, consider the Georgi-Glashow model (\ref{ggl}). With
\begin{eqnarray}
%\label{eq:Tri}
\vec{\Phi} = \rho~\hat{n},
~~~\A_\mu=\hat A_\mu +\vec W_\mu,
\end{eqnarray}
we have the Abelian decomposition,
\bea
&{\cal L}_{GG} =-\dfrac{1}{2} (\partial_\mu \rho)^2
-\dfrac{g^2}{2} {\rho}^2 (\W_\mu)^2-\dfrac{\lambda}{4}
\big(\rho^2 -\dfrac{\mu^2}{\lambda}\big)^2 \nn\\
&-\dfrac{1}{4} {\hat F}_{\mu\nu}^2 
-\dfrac{1}{4}(\D_\mu\W_\nu-\D_\nu\W_\mu)^2 \nn\\
&-\dfrac{g}{2}{\hat F}_{\mu\nu} \cdot (\W_\mu \times \W_\nu)
-\dfrac{g^2}{4} (\W_\mu \times \W_\nu)^2. 
\label{gglag1}
\eea
With this we can Abelianize it gauge independently,
\begin{gather}
{\cal L}_{GG}= -\frac{1}{2} (\partial_\mu \rho)^2
- g^2 {\rho}^2 |W_\mu |^2-\frac{\lambda}{4}\big(\rho^2 
-\frac{\mu^2}{\lambda}\big)^2 \nn\\
- \frac{1}{4} {\cal F}_{\mu\nu}^2
-\frac{1}{2} |\D_\mu W_\nu-\D_\nu W_\mu|^2
+ ig {\cal F}_{\mu\nu} W_\mu^*W_\nu \nn\\
+ \frac{g^2}{4}(W_\mu^* W_\nu - W_\nu^* W_\mu)^2.
\label{gglag2}
\end{gather}
This clearly shows that the theory can be viewed as a 
(non-trivial) Abelian gauge theory which has a charged 
vector field as a source. 

The Abelianized Lagrangian looks very much like the 
Georgi-Glashow Lagrangian written in the unitary gauge. 
But we emphasize that this is the gauge independent 
Abelianization which has the full (quantum) $SU(2)$ 
gauge symmetry.  

Obviously we can apply the same Abelian decomposition 
to the Weinberg-Salam theory
\bea
&{\cal L}=-\dfrac{1}{2}{(\partial_{\mu}\rho)}^2
-\dfrac{\rho^2}{2} {|{\cal \hat D}_{\mu} \xi |}^2
-\dfrac{\lambda}{8}(\rho^2-\rho_0^2)^2 \nn\\
&-\dfrac{1}{4} {\hat F}_{\mu\nu}^2 
-\dfrac{1}{4} G_{\mu\nu}^2
-\dfrac{1}{4}(\D_\mu\W_\nu-\D_\nu\W_\mu)^2 
-\dfrac{g^2}{8}\rho^2 (\W_\mu)^2  \nn\\
&-\dfrac{g}{2}{\hat F}_{\mu\nu} \cdot (\W_\mu \times \W_\nu)
-\dfrac{g^2}{4} (\W_\mu \times \W_\nu)^2,  \nn\\
&{\cal \hat D}_\mu=\pro_\mu
-i\dfrac{g}{2} \vec{\tau}\cdot\hat {A}_\mu-i\dfrac{g'}{2}B_\mu.
\label{wslag2}
\eea
Moreover, with
\begin{gather}
\left( \begin{array}{cc} A_\mu^{\rm (em)} \\  Z_{\mu}
\end{array} \right)
= \frac{1}{\sqrt{g^2 + g'^2}} \left(\begin{array}{cc} g & g' \\
-g' & g
\end{array} \right)
\left( \begin{array}{cc}
B_{\mu} \\ {\cal A}_{\mu}
\end{array} \right), 
\label{mixing}
\end{gather}
we can Abelianize it gauge independently
\begin{gather}
{\cal L}= -\frac{1}{2}(\partial_\mu \rho)^2 
-\frac{\lambda}{8}\big(\rho^2-\rho_0^2 \big)^2 \nn\\
-\frac{1}{4} {F_{\mu\nu}^{\rm (em)}}^2 
-\frac{1}{4} Z_{\mu\nu}^2-\frac{g^2}{4}\rho^2 |W_\mu|^2
-\frac{g^2+g'^2}{8} \rho^2 Z_\mu^2 \nn\\
-\frac{1}{2}|(D_\mu^{\rm (em)} W_\nu - D_\nu^{\rm (em)} W_\mu)
+ ie \frac{g}{g'} (Z_\mu W_\nu - Z_\nu W_\mu)|^2  \nn\\
+ie F_{\mu\nu}^{\rm (em)} W_\mu^* W_\nu
+ie \frac{g}{g'}  Z_{\mu\nu} W_\mu^* W_\nu \nn\\
+ \frac{g^2}{4}(W_\mu^* W_\nu - W_\nu^* W_\mu)^2,
\label{wslag3}
\end{gather}
where $D_\mu^{\rm (em)}=\partial_\mu+ieA_\mu^{\rm (em)}$.
Again we emphasize that this is not the Weinberg-Salam Lagrangian 
in the unitary gauge. This is the gauge independent Abelianization 
which has the extra quantum (passive) non-Abelian gauge degrees of 
freedom. This can easily be understood comparing (\ref{mixing}) 
with (\ref{wein}). Certainly (\ref{mixing}) is gauge independent, while
(\ref{wein}) applies to the unitary gauge.

This provides us important piece of information. In the absence 
of the electromagnetic interaction (i.e., with $A_\mu^{\rm (em)}
= W_\mu = 0$) the Weinberg-Salam model describes a spontaneously 
broken $U(1)_Z$ gauge theory,
\begin{gather}
{\cal L} = -\frac{1}{2}(\partial_\mu \rho)^2
-\frac{\lambda}{8}\big(\rho^2-\rho_0^2\big)^2  \nn\\
-\frac{1}{4} Z_{\mu\nu}^2-\frac{g^2+g'^2}{8} \rho^2 Z_\mu^2,
\end{gather}
which is nothing but the Ginsburg-Landau theory of superconductivity.
Furthermore, here $M_H$ and $M_Z$ corresponds to the coherence length
(of the Higgs field) and the penetration length (of the magnetic
field made of $Z$-field). So, when $M_H > M_Z$ (or $M_H < M_Z$),
the theory describes a type II (or type I) superconductivity,
which is well known to admit the Abrikosov-Nielsen-Olesen
vortex solution. This confirms the existence of Nambu's string 
in Weinberg-Salam model. What Nambu showed was that he could make 
the string finite by attaching the fractionally charged monopole
anti-monopole pair to this string \cite{nambu}.

\section{Comparison with Julia-Zee Dyon}

The Cho-Maison dyon looks very much like the well-known
Julia-Zee dyon in the Georgi-Glashow model. Both can be 
viewed as the Wu-Yang monopole dressed by the weak boson(s). 
However, there is a crucial difference. The the Julia-Zee 
dyon is completely regular and has a finite energy, while 
the Cho-Maison dyon has a point singularity at the center 
which makes the energy infinite. 

So, to regularize the Cho-Maison dyon it is important 
to understand the difference between the two dyons.
To do that notice that, in the absence of the 
$Z$-boson, (\ref{wslag3}) reduces to
\bea
&{\cal L}= -\dfrac{1}{2}(\partial_\mu \rho)^2 
-\dfrac{\lambda}{8}\big(\rho^2-\rho_0^2\big)^2
-\dfrac{g^2}{4}\rho^2 |W_\mu|^2 \nn\\
&-\dfrac{1}{4} {F_{\mu\nu}^{\rm (em)}}^2 
-\dfrac{1}{2}|D_\mu^{\rm (em)} W_\nu-D_\nu^{\rm (em)} W_\mu|^2 \nn\\
&+ie F_{\mu\nu}^{\rm (em)} W_\mu^* W_\nu 
+ \dfrac{g^2}{4}(W_\mu^* W_\nu - W_\nu^* W_\mu)^2.
\label{wslag4}
\eea
This should be compared with (\ref{gglag2}), which shows that
the two theories have exactly the same type of interaction in 
the absence of the $Z$-boson, if we identify $\cal F_\mn$ in 
(\ref{gglag2}) with $F_{\mu\nu}^{\rm (em)}$ in (\ref{wslag4}). The 
only difference is the coupling strengths of the $W$-boson 
quartic self-interaction and Higgs interaction of $W$-boson 
(responsible for the Higgs mechanism). This difference, of 
course, originates from the fact that the Weinberg-Salam model 
has two gauge coupling constants, while the Georgi-Glashow model 
has only one. 

This tells that, in spite of the fact that the Cho-Maison dyon 
has infinite energy, it is not much different from the Julia-Zee 
dyon. To amplify this point notice that the spherically symmetric 
ansatz of the Julia-Zee dyon
\begin{gather}
\vec \Phi=\rho(r)~\hat r, 
~~~\hat A_\mu=\frac{1}{g}A(r)\partial_\mu t~\hat{r}
- \frac{1}{g}\hat{r}\times\partial_\mu \hat{r} \nn\\
\W_\mu= \frac{1}{g}f(r) \hat{r}\times\partial_\mu \hat{r} ,
\label{ggdans}
\end{gather}
can be written in the Abelian formalism as
\begin{gather}
\rho = \rho(r), 
~~~W_{\mu} = \frac{i}{g}\frac{f(r)}{\sqrt2}e^{i\varphi}
(\partial_\mu \theta +i \sin\theta \partial_\mu \varphi), \nn\\
{\cal A}_{\mu} = \frac{1}{g}A(r) \partial_{\mu}t 
-\frac{1}{g}(1-\cos\theta) \partial_{\mu} \varphi.
\end{gather}
In the absence of the $Z$-boson this is identical to the 
ansatz (\ref{ans2}).

With the ansatz we have the following equation for the dyon 
\begin{gather}
\ddot{\rho}+\frac{2}{r}\dot{\rho} - 2\frac{f^2}{r^2}\rho 
=\lambda \big(\rho^2 - \frac{\mu^2}{\lambda} \big)\rho, 
\nonumber \\
\ddot{f}- \frac{f^2-1}{r^2}f=(g^2\rho^2-A^2)f, 
\nonumber\\
\ddot{A}+\frac{2}{r}\dot{A} -2\frac{f^2}{r^2} A=0.
\label{ggdeq}
\end{gather}
This should be compared to the equation of motion (\ref{cmeq})
for the Cho-Maison dyon. They are not much different. 

With the boundary condition
\begin{gather}
\rho(0)=0, \quad  f(0)=1, \quad A(0)=0, \nn\\
\rho(\infty)=\bar \rho_0=\sqrt{\mu^2/\lambda},~~f(\infty)=0,
~~A(\infty)=A_0,
\label{ggbc}
\end{gather}
one can integrate (\ref{ggdeq}) and obtain the Julia-Zee 
dyon which has a finite energy. Notice that the boundary condition 
$A(0)=0$ and $f(0)=1$ is crucial to make the solutions regular 
at the origin. This confirms that the Julia-Zee dyon is nothing 
but the Abelian monopole regularized by $\rho$ and $W_\mu$, where 
the charged vector field adds an extra electric charge to the 
monopole. Again it must be clear from (\ref{ggdeq}) that, 
for a given magnetic charge, there are always two dyons with 
opposite electric charges. 

Moreover, for the monopole (and anti-monopole) solution 
with $A=0$, the equation reduces to the following 
Bogomol'nyi-Prasad-Sommerfield equation in the limit 
$\lambda=0$
\begin{gather}
\dot{\rho}\pm \frac{1}{gr^2}(f^2-1)=0,
~~~\dot{f} \pm g \rho f=0. 
\label{pseq}
\end{gather}
This has the analytic solution
\begin{gather}
\rho= \bar \rho_0\coth(g\bar \rho_0 r)-\dfrac{1}{er},
~~~f= \dfrac{g\bar \rho_0 r}{\sinh(g \bar \rho_0 r)},
\end{gather}
which describes the Prasad-Sommerfield monopole \cite{prasad}.

Of course, the Cho-Maison dyon has a non-trivial dressing of 
the $Z$-boson which is absent in the Julia-Zee dyon. But notice 
that the $Z$-boson plays no role in the Cho-Maison monopole. 
This confirms that the Cho-Maison monopole and the `tHooft-Polyakov 
monopole are not so different, so that the Cho-Maison monopole 
could be modified to have finite energy.

For the anti-dyon we can have the following ansatz 
\begin{gather}
\vec \Phi=\rho(r)~\hr', 
~~~\hat A_\mu'=-\frac{1}{g}A(r)\partial_\mu t~\hr'
- \frac{1}{g}\hr' \times\partial_\mu \hr' \nn\\
\W_\mu'= \frac{1}{g}f(r)~\hr' \times\partial_\mu \hr', \nn\\
\hr'=(\sin \theta \cos \phi,-\sin \theta \sin \phi,\cos \theta),
\label{antiggd}
\end{gather}
or equivalently
\begin{gather}
\rho' = \rho(r), 
~~~W_{\mu} = \frac{i}{g}\frac{f(r)}{\sqrt2}e^{-i\varphi}
(\partial_\mu \theta -i \sin\theta \partial_\mu \varphi), \nn\\
{\cal A}'_{\mu} = -\frac{1}{g}A(r) \partial_{\mu}t 
+\frac{1}{g}(1-\cos\theta) \partial_{\mu} \varphi.
\end{gather}
This ansatz looks different from the popular ansatz described
by $\vec \Phi=-\rho(r)~\hr$, but we can easily show that they 
are gauge equivalent. With this we have exactly the same equation 
(\ref{ggdeq}) for the anti-dyon, which assures that the theory 
has both dyon and anti-dyon. 

\section{Ultraviolet Regularization of Cho-Maison Dyon}

Since the Cho-Maison dyon is the only dyon in the standard
model, it is impossible to regularize it within the model. 
However, the Weinberg-Salam model is the ``bare" theory 
which should change to the ``effective" theory after the 
quantum correction, and the ``real" electroweak dyon must 
be the solution of such theory. So we may hope that the 
quantum correction could regularize the Cho-Maison dyon. 

The importance of the quantum correction in classical solutions 
is best understood in QCD. The ``bare" QCD Lagrangian has no 
confinement, so that the classical solutions of the bare QCD 
can never describe the quarkonium or hadronic bound states. 
Only the effective theory can do. 

To see how the quantum modification could make the energy of 
the Cho-Maison monopole finite, notice that after the quantum 
correction the coupling constants change to the scale dependent 
running couplings. So, if this quantum correction makes $1/g'^2$ 
in $E_0$ in (\ref{cme}) vanishing in the short distance limit, the 
Cho-Maison monopole could have finite energy. 

To do that consider the following effective Lagrangian which 
has the non-canonical kinetic term for the $U(1)_Y$ gauge field 
\begin{gather}
{\cal L}_{eff} = -|{\cal D} _\mu \phi|^2
-\frac{\lambda}{2} \Big(\phi^2 -\frac{\mu^2}{\lambda}\Big)^2 
-\frac{1}{4} \vec F_\mn^2  \nn\\ 
-\frac{1}{4} \epsilon(|\phi|^2 ) G_\mn^2,
\label{effl}
\end{gather}
where $\epsilon(|\phi|^2)$ is a positive dimensionless function 
of the Higgs doublet which approaches to one asymptotically. 
Clearly $\epsilon$ modifies the permittivity of the $U(1)_Y$ 
gauge field, but the effective action still retains the 
$SU(2)\times U(1)_Y$ gauge symmetry. Moreover, when 
$\epsilon \rightarrow 1$ asymptotically, the effective 
action reproduces the standard model. 

This type of effective theory which has the field dependent 
permittivity naturally appears in the non-linear electrodynamics
and higher-dimensional unified theory, and has been studied 
intensively in cosmology to explain the late-time accelerated 
expansion \cite{prd87,prl92,babi}.

\begin{figure}
\includegraphics[height=4cm, width=8cm]{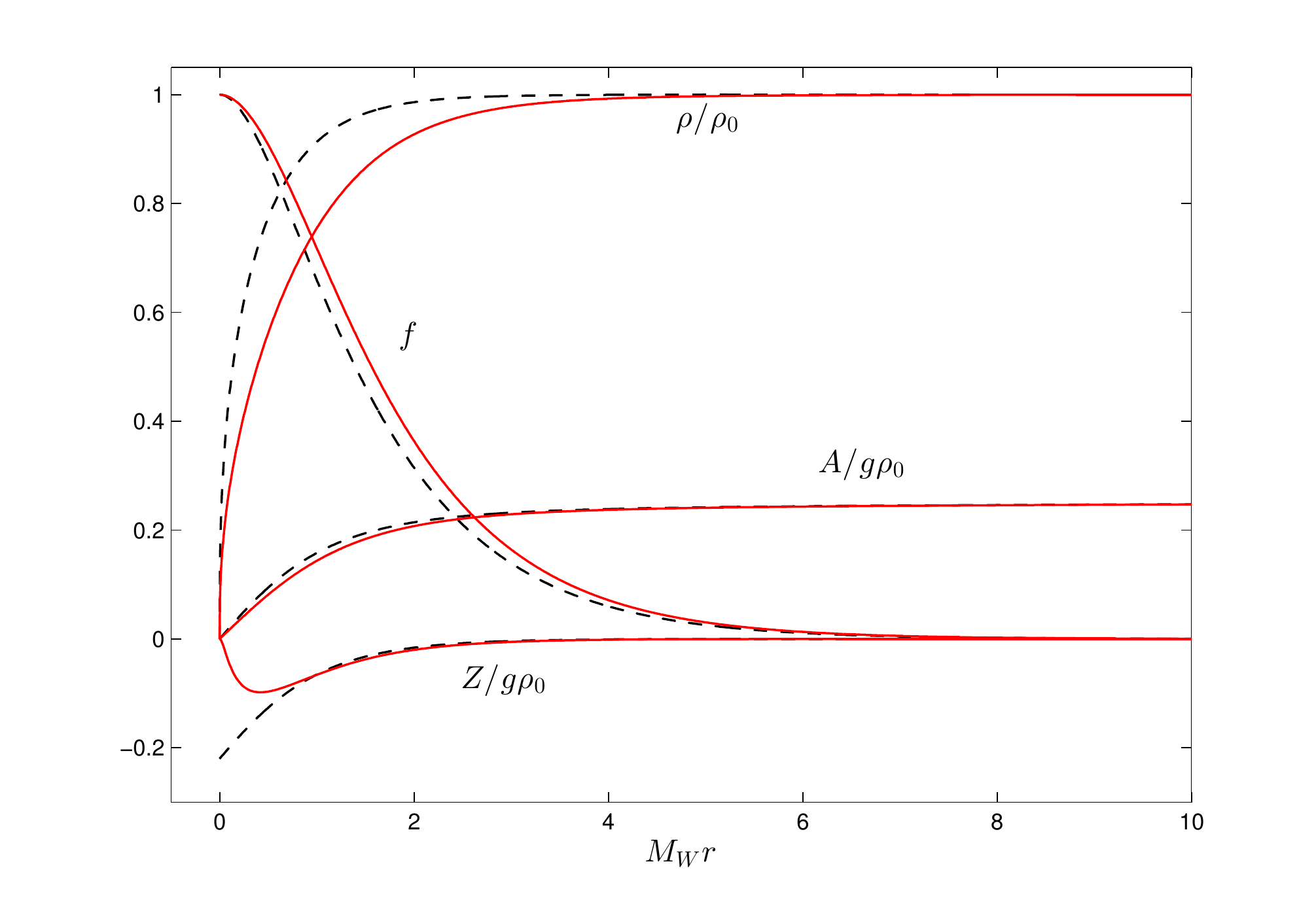}
\caption{\label{fcdyon} The finite energy electroweak dyon 
solution obtained from the effective Lagrangian (\ref{effl}). 
The solid line represents the finite energy dyon 
and dotted line represents the Cho-Maison dyon, where 
$Z=A-B$ and we have chosen $f(0)=1$ and $A(\infty)=M_W/2$.}
\label{fig2}
\end{figure}

From (\ref{effl}) we have the equations for $\rho$ and $B_\mu$ 
\begin{gather}
\partial^2 \rho =|{\cal D}_\mu \xi|^2 \rho
+\frac{\lambda}{2}(\rho^2 - \rho_0^2) \rho 
+\frac{1}{2} \epsilon' \rho G_{\mu\nu}^2,  \nn \\ 
\partial_\mu G_\mn= i \frac{g'}{2 \epsilon} \rho^2 
[\xi^\dagger {\cal D}_\nu \xi - ({\cal D}_\nu\xi)^\dagger \xi] 
-\frac{\partial_\mu \epsilon}{\epsilon} G_\mn, 
\label{meq}
\end{gather}
where $\epsilon' = d\epsilon/d{\rho^2}$. This changes the dyon 
equation (\ref{cmeq}) to 
\begin{gather}
\ddot{\rho} + \frac{2}{r}\dot{\rho}-\frac{f^2}{2r^2}\rho 
=-\frac{1}{4} (A-B)^2 \rho 
+\frac{\lambda}{2} (\rho^2- \rho_0^2) \rho \nn\\
+ \frac{\epsilon'}{g'^2}\Big(\frac{1}{r^4}-\dot{B}^2 \Big) \rho,  \nn\\
\ddot{f}-\frac{f^2-1}{r^2}f=\big(\frac{g^2}{4}\rho^2
- A^2\big)f, \nn\\
\ddot{A}+\frac{2}{r}\dot{A}-\frac{2f^2}{r^2}A
=\frac{g^2}{4}\rho^2(A-B), \nn \\
\ddot{B} + 2\big(\frac{1}{r}+
\frac{\epsilon'}{\epsilon} \rho \dot{\rho} \big) \dot{B}  
=-\frac{g'^2}{4 \epsilon} \rho^2 (A-B).
\end{gather}
This tells that effectively $\epsilon$ changes the 
$U(1)_Y$ gauge coupling $g'$ to the ``running" coupling 
$\bar g'=g' /\sqrt{\epsilon}$. This is because with 
the rescaling of $B_\mu$ to $B_\mu/g'$, $g'$ changes 
to $g' /\sqrt{\epsilon}$. So, by making $\bar g'$ 
infinite (requiring $\epsilon$ vanishing) at the 
origin, we can regularize the Cho-Maison monopole.

From the equations of motion we find that we need the
following condition near the origin to make the monopole 
energy finite
\begin{gather}
\epsilon \simeq \Big(\frac{\rho}{\rho_0}\Big)^n,
~~~n > 4+2\sqrt 3 \simeq 7.46.
\end{gather}
With $n=8$ we have
\bea
\rho(r) \simeq  r^\delta,
~~~\delta={\frac{\sqrt{3}-1}{2}},
\eea
near the origin, and have the finite energy dyon solution 
shown in Fig. \ref{fcdyon}. It is really remarkable 
that the regularized solutions look very much like the 
Cho-Maison solutions, except that for the finite energy 
dyon solution $Z(0)$ becomes zero. This confirms that 
the ultraviolet regularization of the Cho-Maison 
monopole can indeed be possible. 

\begin{figure}
\includegraphics[width=8cm, height=4cm]{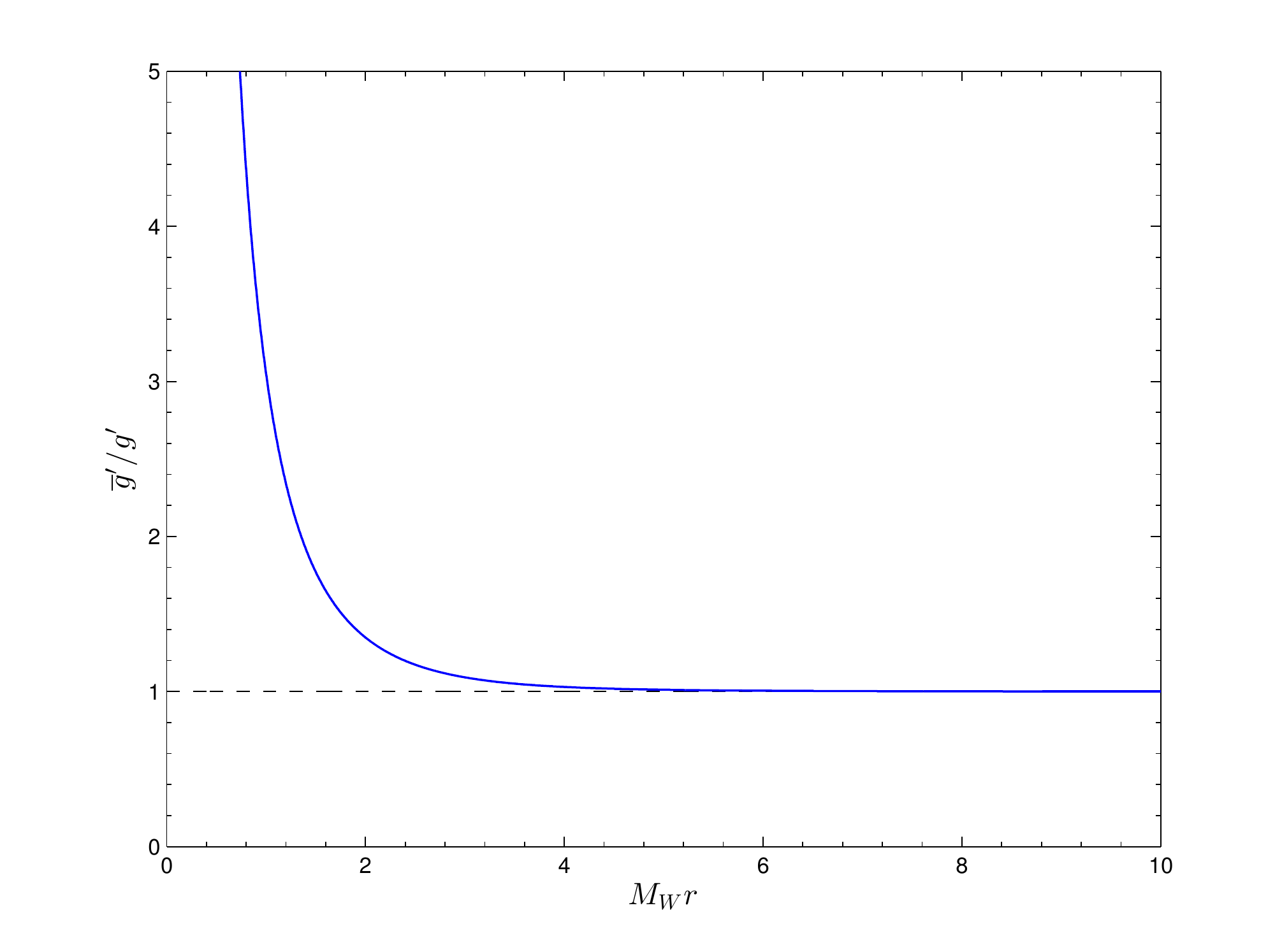}
\caption{\label{effg} The running coupling $\bar g'$ of $U(1)_Y$
gauge field induced by the effective Lagrangian (\ref{effl}).} 
\end{figure}

\begin{figure}
\includegraphics[width=8cm, height=4cm]{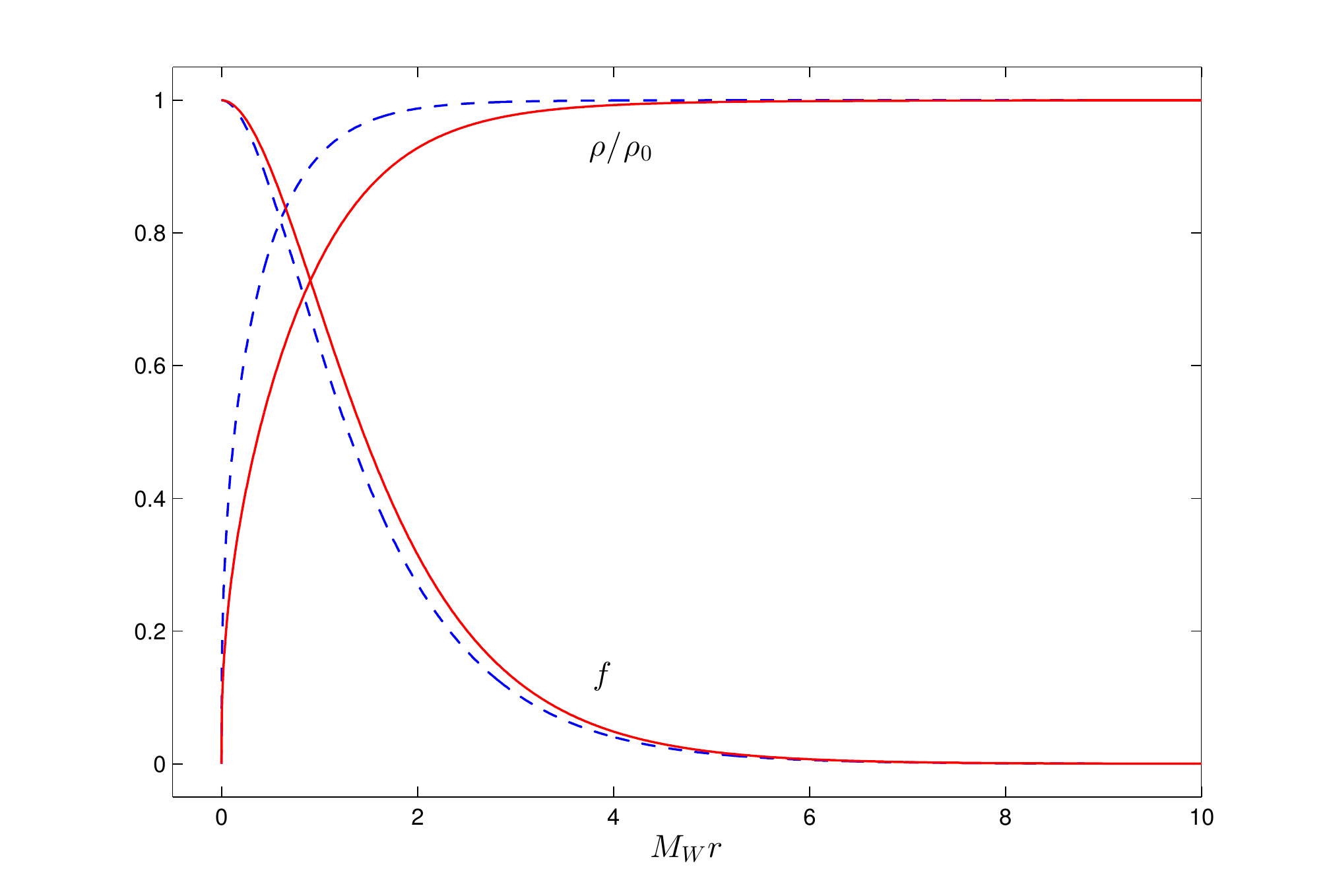}
\caption{\label{fcmono}The finite energy electroweak monopole 
solution obtained from the effective Lagrangian (\ref{modl}). 
The solid line (red) represents the regularized monopole and 
the dotted (blue) line represents the Cho-Maison monopole.}
\end{figure}

As expected with $n=8$ the running coupling $\bar g'$ becomes
divergent at the origin, which makes the energy contribution 
from the $U(1)_Y$ gauge field finite. The scale dependence of 
the running coupling is shown in Fig. \ref{effg}. With $A=B=0$ 
we can estimate the monopole energy to be
\begin{gather}
E \simeq 0.65 \times \frac{4\pi}{e^2} M_W \simeq 7.19 ~{\rm TeV}.
\end{gather}
This tells that the estimate of the monopole energy based on 
the scaling argument is reliable. The finite energy monopole 
solution is shown in Fig.~\ref{fcmono}. 

There is another way to regularize the Cho-Maison monopole.
Suppose we have the following ultraviolet modification of 
(\ref{wslag2}) from the quantum correction
\begin{gather}
\delta {\cal L}=ie \alpha F_{\mu\nu}^{\rm (em)} W_\mu^* W_\nu 
+\beta \frac{g^2}{4}(W_\mu^*W_\nu-W_\nu^*W_\mu)^2  \nn\\
-\gamma \frac{g^2}{4} \rho^2 |W_\mu|^2.  
\end{gather}
where $\alpha,~\beta,~\gamma$ are the scale dependent parameters 
which vanish asymptotically (and modify the theory only at short distance). 
With this we have the modified Weinberg-Salam Lagrangian
\begin{gather}
{\cal L}'= -\frac{1}{2}(\partial_\mu \rho)^2 
-\frac{\lambda}{8}\big(\rho^2-\rho_0^2\big)^2  
-\frac{1}{4} {F_{\mu\nu}^{\rm (em)}}^2
-\frac{1}{4} Z_{\mu\nu}^2  \nn\\
-\frac{1}{2}\big|(D_\mu^{\rm (em)} W_\nu - D_\nu^{\rm (em)} W_\mu)
+ie \frac{g}{g'}(Z_\mu W_\nu - Z_\nu W_\mu)\big|^2 \nn \\
+ie(1+\alpha) F_{\mu\nu}^{\rm (em)} W_\mu^* W_\nu 
+ie \frac{g}{g'} Z_{\mu\nu} W_\mu^* W_\nu \nn\\
+(1+\beta)\frac{g^2}{4}(W_\mu^* W_\nu -W_\nu^* W_\mu)^2 \nn\\
-(1+\gamma) \frac{g^2}{4}\rho^2 |W_\mu|^2
-\frac{g^2+g'^2}{8} \rho^2 Z_\mu^2.
\label{modl}
\end{gather}
Of course, this modification is supposed to hold only in the 
short distance, so that asymptotically  $\alpha,~\beta,~\gamma$
should vanish to make sure that ${\cal L}'$ reduces to the 
standard model. But we will treat them as constants, partly 
because it is difficult to make them scale dependent, but
mainly because asymptotically the boundary condition automatically 
makes them irrelevant and assures the solution to converge to 
the Cho-Maison solution.

To understand the physical meaning of (\ref{modl}) notice 
that in the absence of the $Z$-boson the above Lagrangian
reduces to the Georgi-Glashow Lagrangian where the $W$-boson 
has an extra ``anomalous" magnetic moment $\alpha$ when 
$(1+\beta)=e^2/g^2$ and $(1+\gamma)=4e^2/g^2$, if we 
identify the coupling constant $g$ in the Georgi-Glashow model 
with the electromagnetic coupling constant $e$. Moreover, the 
ansatz (\ref{ans1}) can be written as
\bea
&\A_\mu=\hat A_\mu^{\rm (em)} +\vec W_\mu, \nn\\
&\hat A_\mu^{\rm (em)}=e \big[\dfrac{1}{g^2}A(r)
+\dfrac{1}{g'^2}B(r)\big] \pro_\mu t ~\hat r  
-\dfrac1e \hat r\times \pro_\mu \hat r , \nn\\
&\vec W_\mu=\dfrac{f(r)}{g} \hat r\times \pro_\mu \hat r, \nn\\
&Z_{\mu} = \dfrac{e}{gg'}\big(A(r)-B(r)\big) \partial_{\mu}t.
\label{ans3}
\eea
This shows that, for the monopole (i.e., for $A=B=0$) 
the ansatz becomes formally identical to (\ref{ggdans}) 
if $\vec W_\mu$ is rescaled by a factor $g/e$. This tells 
that, as far as the monopole solution is concerned, in 
the absence of the $Z$-boson the Weinberg-Salam model and 
Georgi-Glashow model are not so different.

With (\ref{modl}) the energy of the dyon is given by 
\begin{gather}
\hat E =\hat E_0 +\hat E_1, \nn\\
\hat E_0 =\frac{2\pi}{g^2}\int_0^\infty
\frac{dr}{r^2}\Big\{\frac{g^2}{g'^2}+1 -2(1+\alpha) f^2
+(1+\beta)f^4 \Big\} \nn\\
=\frac{2\pi}{g^2}\int_0^\infty
\frac{dr}{r^2}\Big\{\frac{g^2}{e^2}-\frac{(1+\alpha)^2}{1+\beta} 
+(1+\beta)\big(f^2-\frac{1+\alpha}{1+\beta}\big)^2 \Big\}, \nn\\
\hat E_1 =\frac{4\pi}{g^2} \int_0^\infty dr 
\bigg\{\frac{g^2}{2}(r\dot\rho)^2 
+\frac{\lambda g^2r^2}{8}\big(\rho^2-\rho_0^2 \big)^2 \nn\\
+\dot f^2 +\frac{1}{2}(r\dot A)^2
+\frac{g^2}{2g'^2}(r\dot B)^2 
+(1+\gamma) \frac{g^2}{4} f^2\rho^2  \nn\\
+\frac{g^2r^2}{8} (B-A)^2 \rho^2 +f^2 A^2 \bigg\}.
\label{energy_2}
\end{gather}
Notice that $\hat E_1$ remains finite with the modification, 
and $\gamma$ plays no role to make the monopole energy 
finite. 

\begin{figure}
\includegraphics[height=4cm, width=8cm]{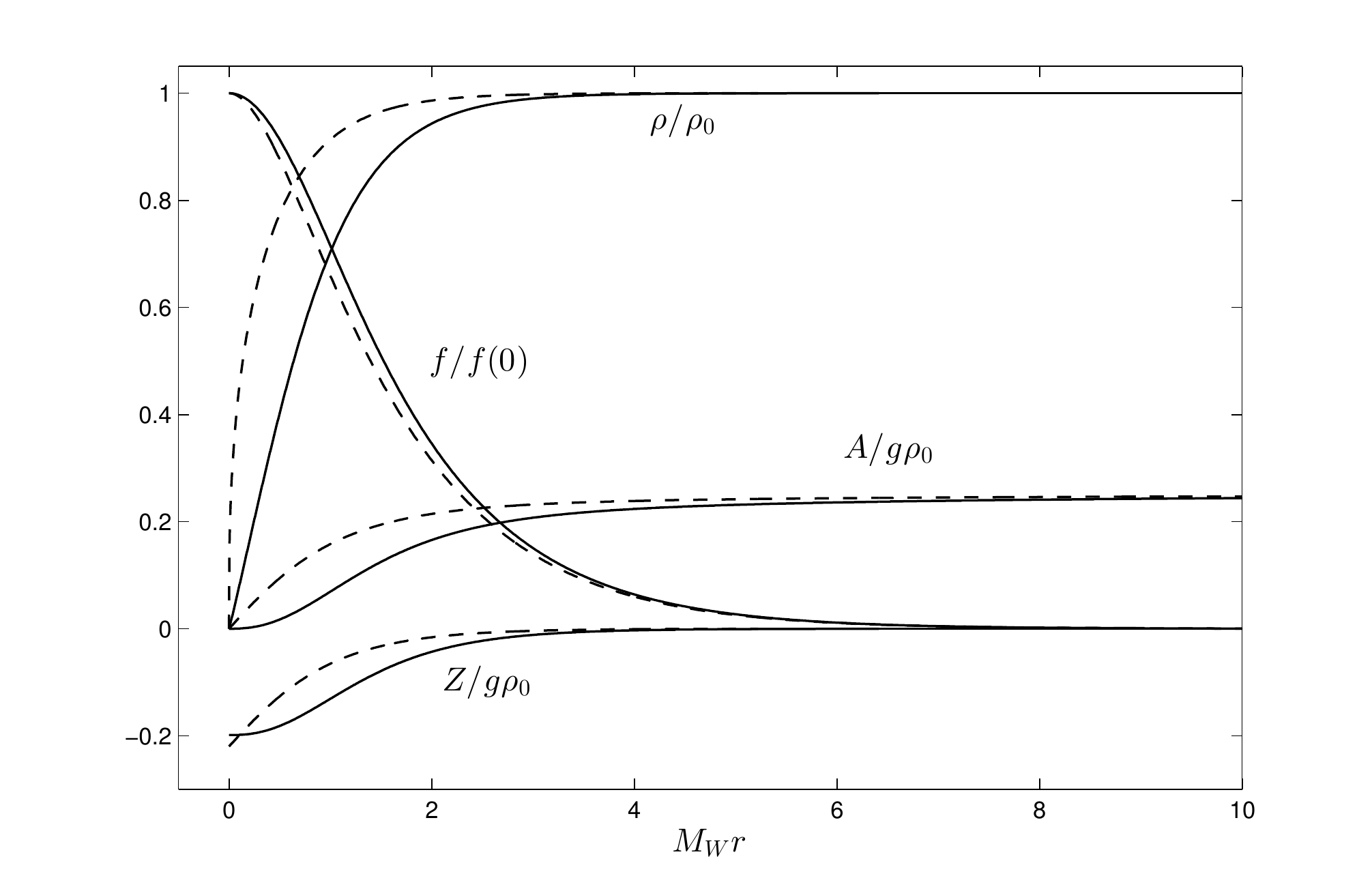}
\caption{\label{fecdyon} The finite energy electroweak dyon 
solution obtained from the modified Lagrangian (\ref{modl}). 
The solid line represents the finite energy dyon and dotted 
line represents the Cho-Maison dyon.}
\label{fig2}
\end{figure}

To make $\hat E_0$ finite we must have
\begin{gather}
1+\alpha=\dfrac1{f(0)^2} \dfrac{g^2}{e^2},
~~~1+\beta=\dfrac1{f(0)^4} \dfrac{g^2}{e^2},
\label{fecon}
\end{gather}
so that the constants $\alpha$ and $\beta$ are fixed by $f(0)$. 
With this the equation of motion is given by
\begin{gather}
\ddot \rho+\frac{2}{r}\dot\rho-\frac{(1+\gamma)f^2}{2r^2}\rho
=-\frac{1}{4}(A-B)^2 \rho
+\frac{\lambda}{2}\big(\rho^2-\rho_0^2 \big)\rho, \nn \\
\ddot f -\frac{(1+\alpha)}{r^2}\Big( \dfrac{f^2}{f^2(0)}-1 \Big) f
=\Big( (1+\gamma)\frac{g^2}{4}\rho^2-A^2 \Big) f, \nn\\
\ddot A +\frac{2}{r} \dot A - \frac{2f^2}{r^2} A
=\frac{g^2}{4}(A-B)\rho^2,  \nn \\
\ddot B+\frac{2}{r}\dot B =-\frac{g'^2}{4}(A-B) \rho^2 . 
\label{eqm3} 
\end{gather}
The solution has the following behavior near the origin,
\bea
&\rho \simeq  \alpha_1 r^{\delta_1},
~~~~\dfrac{f}{f(0)} \simeq 1 + \beta_1 r^{\delta_2}, \nn \\
&A \simeq a_1 r^{\delta_3},~~~~B \simeq b_0 + b_1 r^{\delta_4}, 
\label{origin1}
\eea
where
\bea
&\delta_1 = \dfrac{1}{2}(\sqrt{1+2(1+\gamma)f^2(0)} -1), \nn\\
&\delta_2 = \dfrac{1}{2}(1+\sqrt{8\alpha+9}), 
~~~\delta_3 = \dfrac{1}{2}(\sqrt{1+8f^2(0)} -1),\nn\\
&\delta_4 = \sqrt{1+2f^2(0)} +1. \nn
\eea
Notice that all four deltas are positive (as far as $(1+\alpha)>0$),
so that the four functions are well behaved at the origin.

If we assume $\alpha=\gamma=0$ we have $f(0)=g/e$, 
and we can integrate (\ref{eqm3}) with the boundary condition
\bea
&\rho(0)=0,~~~f(0)=g/e,~~~A(0)=0,~~~B(0)=b_0, \nn\\
&f(\infty)=0,~\rho(\infty)=\rho_0,
~A(\infty)=B(\infty)=A_0.
\label{bc1}
\eea
The finite energy dyon solution is shown in Fig. \ref{fecdyon}. 
It should be emphasized that the solution is an approximate 
solution which is supposed to be valid only near the origin, 
because the constants $\alpha,~\beta,~\gamma$ are supposed 
to vanish asymptotically. But notice that asymptotically the 
solution automatically approaches to the Cho-Maison solution 
even without making them vanish, because we have the same 
boundary condition at the infinity. Again it is remarkable that 
the finite energy solution looks very similar to the Cho-Maison 
solution.  

\begin{figure}
\includegraphics[height=4cm, width=8cm]{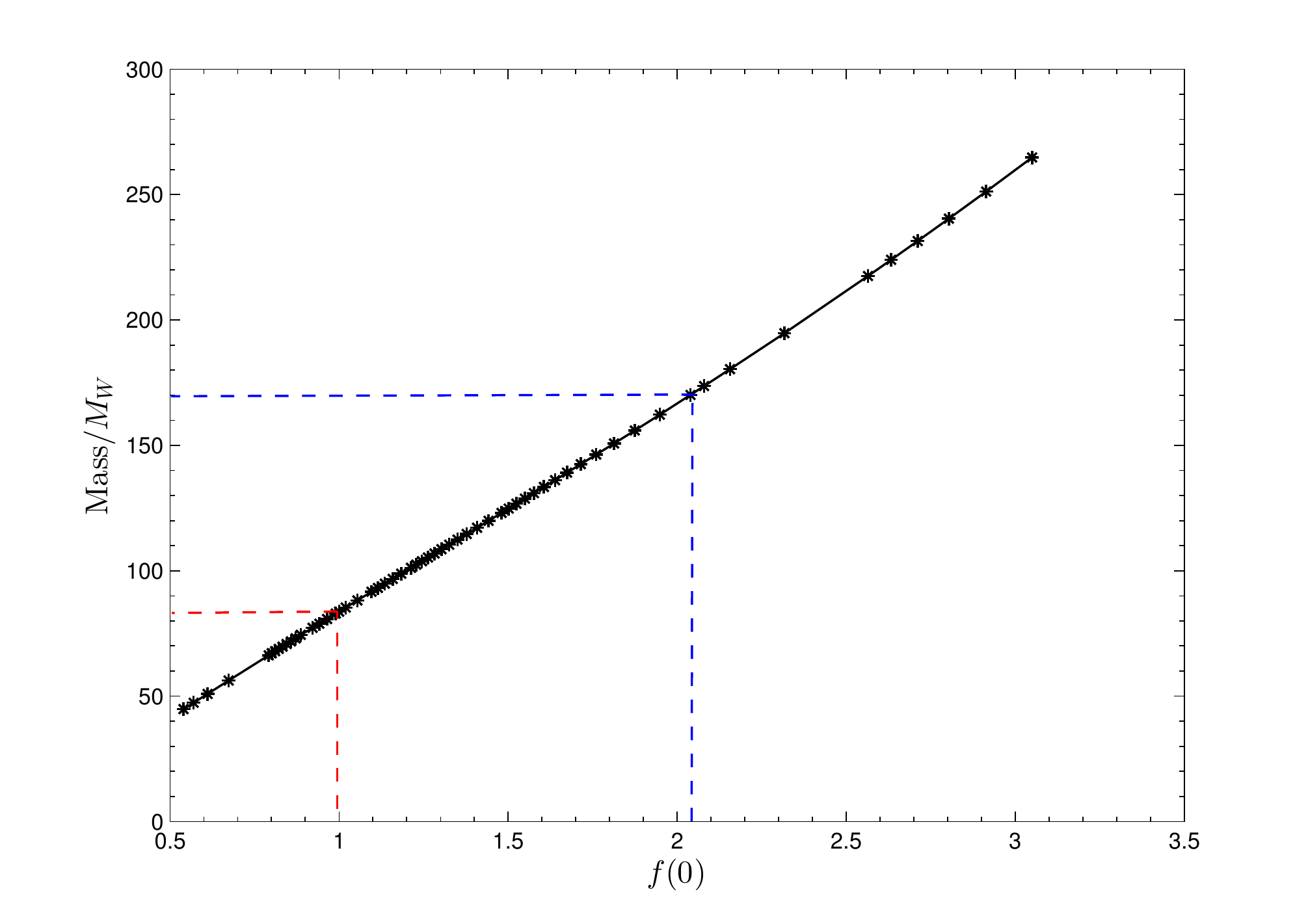}
\caption{\label{edf0} The energy dependence of the electroweak 
monopole on $f(0)$.}
\label{fig3}
\end{figure}

Of course, we can still integrate (\ref{eqm3}) with arbitrary 
$f(0)$ and have a finite energy solution.  The monopole energy for 
$f(0)=1$ and $f(0)=g/e$ (with $\alpha=\gamma=0$)  are given 
by
\begin{eqnarray}
&E(f(0)=1) \simeq 0.61 \times \dfrac{4\pi }{e^2} M_W 
\simeq 6.73~{\rm TeV},  \nn\\
&E(f(0)=\dfrac{g}{e})\simeq 1.27 \times \dfrac{4\pi }{e^2} M_W 
\simeq 13.95~{\rm TeV}.
\end{eqnarray}
In general the energy of dyon depends on $f(0)$, but must be 
of the order of $(4\pi/e^2) M_W$. The energy dependence of 
the monopole on $f(0)$ is shown in Fig. \ref{edf0}. This 
strongly supports our prediction of the monopole mass based 
on the scaling argument.

As we have emphasized, in the absence of the $Z$-boson 
(\ref{modl}) reduces to the Georgi-Glashow theory with
\bea
\alpha = 0, \quad 
1+\beta=\frac{e^2}{g^2}, \quad 
1+\gamma=\frac{4e^2}{g^2}.
\label{cond4}
\eea
In this case (\ref{eqm3}) reduces to the following 
Bogomol'nyi-Prasad-Sommerfield equation in the limit 
$\lambda=0$ \cite{prasad}
\begin{gather}
\dot{\rho}\pm \frac{1}{er^2}\big(\frac{e^2}{g^2}f^2
-1 \big)=0,
~~~\dot{f}\pm e\rho f=0.
\label{self2}
\end{gather}
This has the analytic monopole solution
\begin{gather}
\rho=\rho_0\coth(e\rho_0r)-\frac{1}{er},
~~~f= \frac{g\rho_0 r}{\sinh(e\rho_0r)},
\end{gather}
whose energy is given by the Bogomol'nyi bound
\begin{eqnarray}
E=\sin \theta_{\rm w} \times \frac{8\pi}{e^2} M_{W}
\simeq 5.08~{\rm TeV}.
\end{eqnarray}
From this we can confidently say that the mass of the electroweak 
monopole could be around 4 to 7 TeV.

This confirms that we can regularize the Cho-Maison dyon with
a simple modification of the coupling strengths of the existing 
interactions which could be caused by the quantum correction. 
This provides a most economic way to make the energy of the dyon 
finite without introducing a new interaction in the standard model.

\section{Embedding $U(1)_Y$ to $SU(2)_Y$}

Another way to regularize the Cho-Maison dyon, of course, is
to enlarge $U(1)_Y$ and embed it to another SU(2). This type of 
generalization of the standard model could naturally arise in the 
left-right symmetric grand unification models, in particular in the 
SO(10) grand unification, although this generalization may be too 
simple to be realistic.

To construct the desired solutions we introduce a hypercharged vector 
field $X_\mu$ and a Higgs field $\sigma$, and generalize the Lagrangian
(\ref{wslag2}) adding the following Lagrangian
\bea
&\Delta {\cal L}=-\dfrac{1}{2}|\tilde D_\mu X_\nu-\tilde D_\nu X_\mu|^2
+ig' G_{\mu\nu}X_\mu^* X_\nu \nn\\
&+\dfrac{1}{4}g'^2(X_\mu^* X_\nu -X_\nu^* X_\mu)^2  \nn\\
&-\dfrac{1}{2}(\partial_\mu\sigma)^2 -g'^2\sigma^2 |X_\mu|^2
-\dfrac{\kappa}{4}\big( \sigma^2-\dfrac{m^2}{\kappa}\big)^2,
\label{lag4}
\eea
where $\tilde D_\mu = \partial_\mu +ig' B_\mu$. To understand 
the meaning of it let us introduce a hypercharge $SU(2)$ gauge field
$\vec B_\mu$ and a scalar triplet ${\vec \Phi}$, and consider 
the $SU(2)_Y$ Georgi-Glashow model
\bea
&{\cal L}'=-\dfrac{1}{2}(D_\mu {\vec \Phi})^2
-\dfrac{\kappa}{4}\big({\vec \Phi}^2-\dfrac{m^2}{\kappa}\big)^2
-\dfrac{1}{4} \vec {G}_{\mu\nu}^2.
\eea
Now we can have the Abelian decomposition of this Lagrangian 
with $\vec \Phi=\sigma \n$, and have (identifying $B_\mu$ and $X_\mu$
as the Abelian and valence parts) 
\bea
&{\cal L}'=-\dfrac14 G_\mn^2+ \Delta {\cal L}.
\eea
This clearly shows that Lagrangian (\ref{lag4}) describes nothing but 
the embedding of the hypercharge $U(1)$ to an $SU(2)$ Georgi-Glashow 
model.

Now for a static spherically symmetric ansatz
we choose (\ref{ans1}) and let
\begin{eqnarray}
&\sigma =\sigma(r), \nn \\
&X_\mu =\dfrac{i}{g'}\dfrac{h(r)}{\sqrt{2}}e^{i\varphi} 
(\partial_\mu \theta+i\sin\theta\partial_\mu \varphi).
\label{ansatz3}
\end{eqnarray}
With the spherically symmetric ansatz the equations of motion are 
reduced to
\begin{gather}
\ddot{f}-\frac{f^2-1}{r^2}f
=\big(\dfrac{g^2}{4}\rho^2-A^2\big)f, \nn\\
\ddot{\rho}+\frac{2}{r} \dot{\rho} - \frac{f^2}{2r^2}\rho
=-\frac{1}{4}(A-B)^2\rho + \frac{\lambda}{2}\big(\rho^2 
-\frac{2\mu^2}{\lambda}\big)\rho, \nn\\
\ddot{A} + \frac{2}{r}\dot{A} -\frac{2f^2}{r^2}A 
= \frac{g^2}{4} \rho^2(A-B), \nn\\
\ddot{B} + \frac{2}{r} \dot{B}- \frac{2h^2}{r^2} B
=\frac{g'^2}{4} \rho^2 (B-A),  \nn\\
\ddot h -\frac{h^2-1}{r^2} h =(g'^2\sigma^2-B^2) h, \nn\\
\ddot\sigma +\frac{2}{r}\dot\sigma -\frac{2h^2}{r^2} \sigma
= \kappa\big(\sigma^2-\frac{m^2}{\kappa}\big)\sigma.
\label{eom4}
\end{gather}
Furthermore, the energy of the above configuration is given by
\begin{gather}
E=E_W +E_X,  \nn\\
E_{W}= \frac{4\pi}{g^2}\int_0^\infty dr
\Big\{\dot f^2 +\frac{(f^2-1)^2}{2r^2} 
+\frac{1}{2}(r\dot A)^2  \nn\\
+f^2A^2+\frac{g^2}{2}(r\dot\rho)^2 + \frac{g^2}{4} f^2\rho^2
+\frac{g^2r^2}{8}(A-B)^2\rho^2  \nn\\
+\frac{\lambda g^2r^2}{8}\big(\rho^2-\frac{2\mu^2}{\lambda}\big)^2
\Big\}=\frac{4\pi}{g^2}~C_1~M_W,  \nn\\
E_X=\frac{4\pi}{g'^2}\int_0^\infty dr\Big\{\dot h^2
+\frac{(h^2-1)^2}{2r^2}+\frac{1}{2}(r\dot B)^2  \nn\\
+h^2B^2 +\frac{g'^2}{2}(r\dot\sigma)^2+g'^2 h^2\sigma^2  \nn\\
+\frac{\kappa g'^2r^2}{4}(\sigma^2-\sigma_0^2)^2 \Big\} 
=\frac{4\pi }{g'^2}~C_2~M_X,  
\end{gather}
where $\sigma_0=\sqrt{m^2/\kappa}$, $M_X=g' \sigma_0$, $C_1$ and $C_2$ 
are constants of the order one. The boundary conditions 
for a regular field configuration can be chosen as
\begin{eqnarray}
&f(0)=h(0)=1,~~ A(0)=B(0)=\rho(0)=\sigma(0)=0, \nonumber\\
&f(\infty)=h(\infty)=0,~A(\infty)=A_0,~B(\infty)=B_0,\nonumber\\
&\rho(\infty)=\rho_0,~\sigma(\infty)=\sigma_0.
\label{bound3}
\end{eqnarray}
Notice that this guarantees the analyticity of
the solution everywhere, including the origin.

\begin{figure}
\includegraphics[height=4cm, width=8cm]{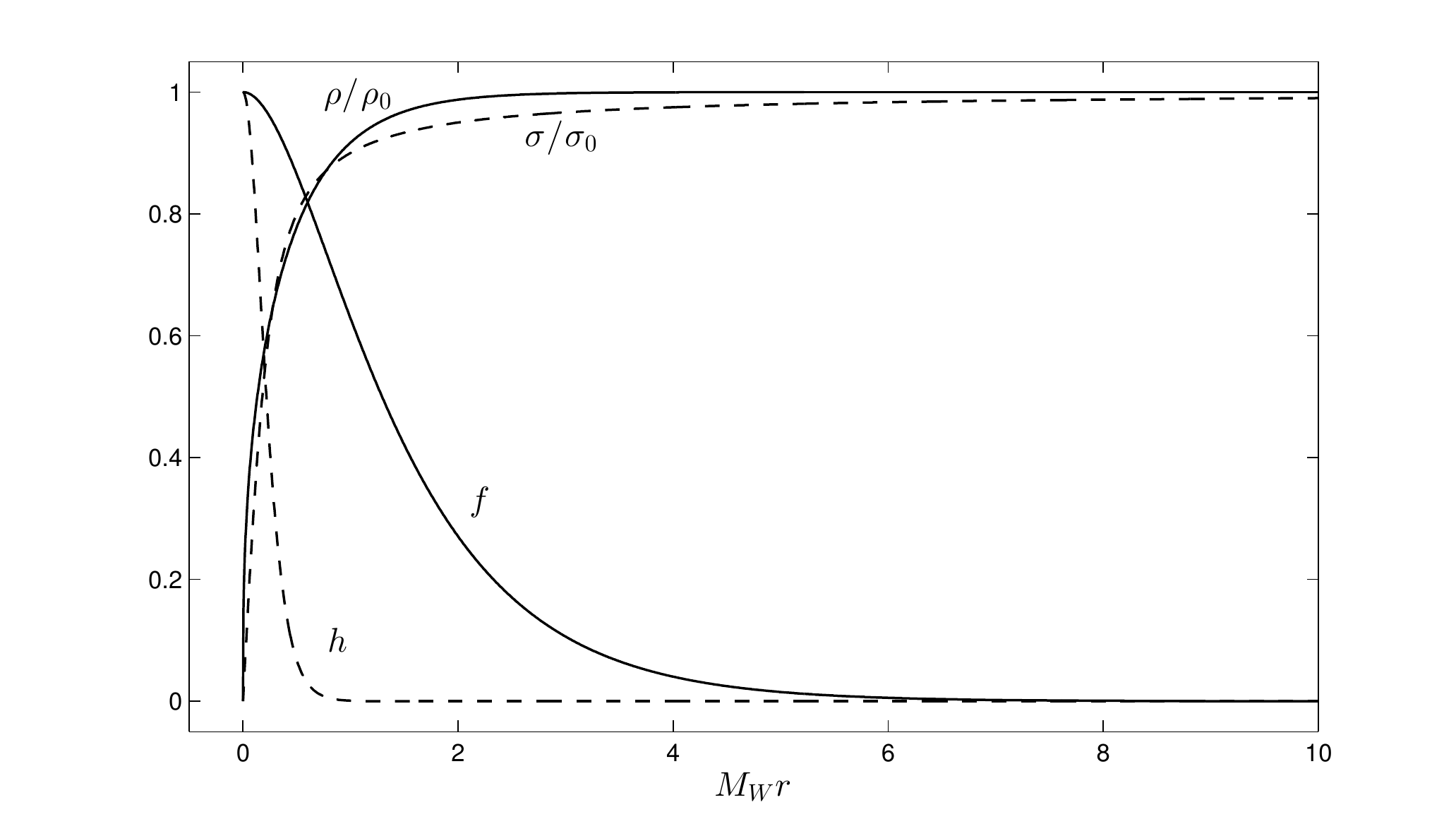}
\caption{The $SU(2)\times SU(2)$ monopole solution with 
$M_H/M_W=1.56$, $M_X=10~M_W$, and $\kappa=0$.}
\label{fig4}
\end{figure}

With the boundary condition (\ref{bound3}) one may try to 
find the desired solution. From the physical point of view 
one could assume $M_X \gg M_W$, where $M_X$ is an intermediate 
scale which lies somewhere between the grand unification scale 
and the electroweak scale. Now, let $A=B=0$ for simplicity. 
Then (\ref{eom4}) decouples to describe two independent systems 
so that the monopole solution has two cores, the one with the 
size $O(1/M_W)$ and the other with the size $O(1/M_X)$. With 
$M_X=10M_W$ we obtain the solution shown in Fig. \ref{fig4}
in the limit $\kappa=0$ and $M_H/M_W=1.56$.

In this limit we find
$C_1=1.53$ and $C_2=1$ so that the  energy of the solution
is given by
\begin{eqnarray}
&E=\dfrac{4\pi}{e^2}\Big( \cos^2\theta_{\rm w}
+0.153~\sin^2\theta_{\rm w}\Big)~M_X  \nn\\
&\simeq 110.17~M_X.
\end{eqnarray}
Clearly the solution describes the Cho-Maison monopole whose singularity 
is regularized by a Prasad-Sommerfield monopole of the size $O(1/M_X)$.

Notice that, even though the energy of the monopole is fixed by 
the intermediate scale, the size of the monopole is determined 
by the electroweak scale. Furthermore from the outside the monopole 
looks exactly the same as the Cho-Maison monopole. Only the inner core 
is regularized by the hypercharged vector field.

\section{Conclusions}

In this paper we have discussed three ways to estimate 
the mass of the electroweak monopole, the dimensional argument, 
the scaling argument, and the ultraviolet regularization of 
the Cho-Maison monopole. As importantly, we have shown that 
the standard model has the anti-dyon as well as the dyon 
solution, so that they can be produced in pairs.

It has generally been believed that the finite energy monopole 
could exist only at the grand unification scale \cite{dokos}. 
But our result tells that the genuine electroweak monopole 
of mass around 4 to 10 TeV could exist. This strongly implies 
that there is an excellent chance that MoEDAL could actually 
detect such monopole in the near future, because the 14 TeV 
LHC upgrade now reaches the monopole-antimonopole pair production 
threshold. But of course, if the mass of the monopole exceeds 
the LHC threshold 7 TeV, we may have to look for the monopole 
from cosmic ray with the ``cosmic" MoEDAL.   

The importance of the electroweak monopole is that it is the 
electroweak generalization of the Dirac monopole, and that it is 
the only realistic monopole which can be produced and detected.
A remarkable aspect of this monopole is that mathematically it 
can be viewed as a hybrid between the Dirac monopole and the 
'tHooft-Polyakov monopole. 

However, there are two crucial differences. First, the magnetic 
charge of the electroweak monopole is two times bigger than that 
of the Dirac's monopole, so that it satisfes the Schwinger quantization 
condition $q_m=4\pi n/e $. This is because the electroweak generalization 
requires us to embed $U(1)_{\rm em}$ to the U(1) subgroup of SU(2), 
which has the period of $4\pi$. So the magnetic charge of the electroweak 
monopole has the unit $4\pi/e$. 

Of course, the finite energy dyon solutions we discussed in 
the above are not the solutions of the ``bare" standard model. 
Nevertheless they tell us how the Cho-Maison dyon could be 
regularized and how the regularized electroweak dyon would 
look like. From the physical point of view there is no doubt 
that the finite energy solutions should be interpreted as 
the regularized Cho-Maison dyons whose mass (and size) is 
fixed by the electroweak scale. 

We emphasize that, unlike the Dirac's monopole which can exist 
only when $U(1)_{\rm em}$ becomes non-trivial, the electroweak 
monopole must exist in the standard model. So, if the standard 
model is correct, we must have the monopole. {\it In this sense, 
the experimental discovery of the electroweak monopole should 
be viewed as the final topological test of the standard model.} 

Clearly the electroweak monopole invites more difficult questions. 
How can we justify the perturbative expansion and the renormalization
in the presence of the monopole? What are the new physical processes 
which can be induced by the monopole? Most importantly, how can we 
construct the quantum field theory of the monopole? 

Moreover, the existence of the finite energy electroweak monopole 
should have important physical implications. In particular, it could 
have important implications in cosmology, because it can be produced 
after inflation. The physical implications of the monopole will be 
discussed in a separate paper \cite{cho}. 

\textbf{Acknowledgments}

The work is supported in part by the National Research
Foundation (2012-002-134) of the Ministry of
Science and Technology and by Konkuk University.


\begin{thebibliography}{99}
\bibitem{LHC} G. Aad et al. (ATLAS Collaboration), Phys. Lett. 
{\bf B716}, 1 (2012); S. Chatrchyan et al. (CMS Collaboration), 
Phys. Lett. {\bf B716}, 30 (2012).
\bibitem{tev} T. Aaltonen et al. (CDF and D0 Collaborations), 
Phys. Rev. Lett. {\bf 109}, 071804 (2012). 
\bibitem{plb97} Y.M. Cho and D. Maison, Phys. Lett. {\bf B391}, 
360 (1997); W.S. Bae and Y.M. Cho, JKPS {\bf 46}, 791 (2005).
\bibitem{yang} Yisong Yang, Proc. Roy. Soc. {\bf A454}, 155 (1998); 
Yisong Yang, {\it Solitons in Field Theory and Nonlinear Analysis} 
(Springer Monographs in Mathematics), p. 322 (Springer-Verlag) 2001.
\bibitem{pin} J. Pinfold, Rad. Meas. {\bf 44}, 834 (2009);
{\it Progress in High-Energy Physics and Nuclear Safety} 
(edited by V. Begun et al.), p. 217 (Springer Science and Business 
Media) 2009; Y.M. Cho and J. Pinfold, Snowmass Whitepaper, 
arXiv hep-ph/1307.8390.
\bibitem{ijmpa14} B. Acharya {\it et al.}  [MoEDAL Collaboration],
Int. J. Mod. Phys. {\bf A29}, 1430050 (2014).
\bibitem{Dirac} P.A.M. Dirac, Phys. Rev. {\bf 74}, 817 (1948).
\bibitem{wu} T.T. Wu and C.N. Yang, in {\it Properties of Matter 
under Unusual Conditions}, edited by H. Mark and S. Fernbach 
(Interscience, New York) 1969; Phys. Rev. {\bf D12}, 3845 (1975).
\bibitem{prl80} Y.M. Cho, Phys. Rev. Lett. {\bf 44}, 1115 (1980); 
Phys. Lett. {\bf B115}, 125 (1982).
\bibitem{Hooft} G. 't Hooft, Nucl. Phys. {\bf B79}, 276 (1974);
A.M. Polyakov, JETP Lett. {\bf 20}, 194 (1974); 
B. Julia and A. Zee, Phys. Rev. {\bf D11}, 2227 (1975).
\bibitem{prasad} M. Prasad and C. Sommerfield, Phys. Rev. Lett. {\bf 35}, 
760 (1975).
\bibitem{dokos} C. Dokos and T. Tomaras, Phys. Rev. {\bf D21}, 2940 (1980).
\bibitem{vach} T. Vachaspati and M. Barriola, Phys. Rev. Lett. {\bf 69},
1867 (1992); M. Barriola, T. Vachaspati, and M. Bucher, Phys. Rev. 
{\bf D50}, 2819 (1994).
\bibitem{nambu} Y. Nambu, Nucl. Phys. {\bf B130}, 505 (1977);
T. Vachaspati, Phys. Rev. Lett. {\bf 68}, 1977 (1992).
\bibitem{vacha} T. Vachaspati, Nucl. Phys. {\bf B439}, 79 (1995).
\bibitem{cab} B. Cabrera, Phys.  Rev. Lett.  {\bf 48}, 1378~(1982). 
\bibitem{Forg} P. Forg\'acs and N.S. Manton, Commun. Math. Phys.
{\bf 72}, 15 (1980).
\bibitem{manton} R.F. Dashen, B. Hasslacher, and A. Neveu, Phys. Rev.
{\bf D10}, 4138 (1974); N.S. Manton, Phys. Rev. {\bf D28}, 2019 (1983);
F. Klinkhammer and N. Manton, Phys. Rev. {\bf D30}, 2212 (1984).
\bibitem{bais} F.A. Bais and R.J. Russell, Phys. Rev. {\bf D11}, 2692
(1975); Y.M. Cho and P.G.O. Freund, Phys. Rev. {\bf D12}, 1711 (1975);
Y.M. Cho and D.H. Park, J. Math. Phys. {\bf 31}, 695 (1990);
P. Breitenlohner, P. Forg\'acs, and D. Maison, Nucl. Phys. {\bf B383},
357 (1992).
\bibitem{prd80} Y.M. Cho, Phys. Rev. {\bf D21}, 1080 (1980).
\bibitem{prl81}Y.M. Cho, Phys. Rev. Lett. {\bf 46}, 302 (1981);
Phys. Rev. {\bf D23}, 2415 (1981); W.S. Bae, Y.M. Cho, and S.W. Kimm, 
Phys. Rev. {\bf D65}, 025005 (2002).
\bibitem{fadd} L. Faddeev and A. Niemi, Phys. Rev. Lett.
{\bf 82}, 1624 (1999); Phys. Lett. {\bf B449}, 214 (1999).
\bibitem{shab}S. Shabanov, Phys. Lett. {\bf B458}, 322 (1999);
{\bf B463}, 263 (1999); H. Gies, Phys. Rev. {\bf D63}, 125023 (2001).
\bibitem{zucc} R. Zucchini, Int. J. Geom. Meth. Mod. Phys. {\bf 1},
813 (2004).  
\bibitem{bpst} A. Belavin, A. Polyakov, A. Schwartz, and Y. Tyupkin,
Phys. Lett. {\bf B59}, 85 (1975); G. 't Hooft, Phys. Rev. Lett. 
{\bf 37}, 8 (1976).
\bibitem{plb79} Y.M. Cho, Phys. Lett. {\bf B81}, 25 (1979).
\bibitem{dewitt} B.S. DeWitt, Phys. Rev. {\bf 162}, 1195 (1967).
\bibitem{prd00} Y.M. Cho, Phys. Rev. {\bf D62}, 074009 (2000);
Y.M. Cho and D.G. Pak, Phys. Rev. {\bf D65}, 074027 (2002); 
Y.M. Cho, M.L. Walker, and D.G. Pak, JHEP {\bf 05}, 073 (2004); 
Y.M. Cho and M.L. Walker, Mod. Phys. Lett. {\bf A19}, 2707 (2004).
\bibitem{prd13} Y.M. Cho, Franklin H. Cho, and J.H. Yoon,  
Phys. Rev. {\bf D87}, 085025 (2013).
\bibitem{kondo} S. Kato, K. Kondo, T. Murakami, A. Shibata, 
T. Shinohara, and S. Ito, Phys. Lett. {\bf B632}, 326 (2006); 
S. Ito, S. Kato, K. Kondo, T. Murakami, A. Shibata, and T. Shinohara, 
Phys. Lett. {\bf B645}, 67 (2007); {\bf B653}, 101 (2007); {\bf B669}, 
107 (2008).
\bibitem{prd87} Y.M. Cho, Phys. Rev. D{\bf 35}, R2628 (1987);
Y.M. Cho, Phys. Lett. {\bf B199}, 358 (1987).
\bibitem{prl92}  Y.M. Cho, Phys. Rev. Lett. {\bf 68}, 3133 (1992); 
Y.M. Cho and J.H. Yoon, Phys. Rev. D{\bf 47}, 3465 (1993).
\bibitem{babi} E. Babichev, Phys. Rev. D{\bf 74}, 085004 (2006). 
\bibitem{cho} Y.M. Cho, Kyoungtae Kimm, and J.H. Yoon, to be published.
\end{thebibliography}
\end{document}